\documentclass[iop,revtex4]{emulateapj}

\usepackage{ulem}
\usepackage{amsmath}
\usepackage{amssymb}
\usepackage{graphicx}
\usepackage{multirow}
\usepackage{lscape}
\usepackage{enumerate}
\usepackage{threeparttable}
\usepackage[breaklinks,colorlinks,citecolor=blue]{hyperref}

\newcommand{\hcn}{H$^{13}$CN }
\newcommand{\hco}{H$^{13}$CO$^+$ }

\slugcomment{180424 version}

\begin{document}

\title{Interactions between Gas Dynamics and Magnetic Fields in the Massive Dense Cores of the DR21 Filament}
\author{Tao-Chung Ching\altaffilmark{1,2,3}, Shih-Ping Lai\altaffilmark{1,4}, Qizhou Zhang\altaffilmark{2}, Josep M.\ Girart\altaffilmark{5,6,2}, Keping Qiu\altaffilmark{7,8}, and Hauyu B.\ Liu\altaffilmark{9}}
\email{chingtaochung@gmail.com}

\altaffiltext{1}{Institute of Astronomy and Department of Physics, National Tsing Hua University, Hsinchu 30013, Taiwan}
\altaffiltext{2}{Harvard-Smithsonian Center for Astrophysics, 60 Garden Street, Cambridge MA 02138, USA}
\altaffiltext{3}{National Astronomical Observatories, Chinese Academy of Sciences, Beijing 100012, China}
\altaffiltext{4}{Institute of Astronomy and Astrophysics, Academia Sinica, P.O. Box 23-141, Taipei 10617, Taiwan}
\altaffiltext{5}{Institut de Ci\`encies de l'Espai (ICE, CSIC), Can Magrans s/n, E-08193 Cerdanyola del Vall\`es, Catalonia}
\altaffiltext{6}{Institut d'Estudis Espacials de Catalunya (IEEC), E-08034 Barcelona, Catalonia}
\altaffiltext{7}{School of Astronomy and Space Science, Nanjing University, 163 Xianlin Avenue, Nanjing 210023, China}
\altaffiltext{8}{Key Laboratory of Modern Astronomy and Astrophysics (Nanjing University), Ministry of Education, Nanjing 210023, China}
\altaffiltext{9}{European Southern Observatory (ESO), Karl-Schwarzschild-Str. 2, D-85748 Garching, Germany}

\begin{abstract}
We report Submillimeter Array molecular line observations in the 345 GHz band of five massive dense cores, Cyg-N38, Cyg-N43, Cyg-N48, Cyg-N51, and Cyg-N53 in the DR21 filament.
The molecular line data reveal several dynamical features of the cores: 
(1) prominent outflows in all cores seen in the CO and SiO lines,
(2) significant velocity gradients in Cyg-N43 and Cyg-N48 seen in the \hcn and \hco lines suggesting 0.1-pc-scale rotational motions, 
and (3) possible infalls in Cyg-N48 found in the SiO and SO lines. 
Comparing the molecular line data and our dust polarization data in \citet{2017Ching}, we find that the gradients of line-of-sight velocities appear to be randomly oriented relative to the plane-of-sky magnetic fields. Our simulations suggest that this random alignment implies parallel or random alignment between the velocity gradients and magnetic fields in the three dimensional space.
The linewidths of \hcn emission are consistently wider than those of \hco emission in the 3$\arcsec$--10$\arcsec$ detectable scales, which can be explained by the existence of ambipolar diffusion with maximum plane-of-the-sky magnetic field strengths of 1.9 mG and 5.1 mG in Cyg-N38 and Cyg-N48, respectively.
Our results suggest that the gas dynamics may distort the magnetic fields of the cores of into complex structures and ambipolar diffusion could be important in dissipating the magnetic energies of the cores.
\end{abstract}


\section{Introduction}
Massive dense cores (0.1--0.2 pc, $\ga$ 100 M$_{\sun}$, $\sim$ 10$^6$ cm$^{-3}$) are the birth places of high-mass stars together with a cluster of low-mass stars \citep{2007Motte,2017Motte}. Observations have revealed that massive dense cores are dynamical with motions including rotations, infalls, and outflows \citep{1997ZH,1998Zhang,2011bCsengeri,2013Beuther,2014Zhang} and magnetized with magnetic field strengths of 0.1 to few mG \citep{2001Lai,2008Falgarone,2009Girart,2009Tang,2013Girart,2013Qiu,2014Frau,2016Houde,2017Ching}.

As the approximate equal partition of the kinematic and magnetic energies implies important roles of both gas dynamics and magnetic fields in the evolution of massive dense cores \citep{2013Girart,2013Qiu,2014Frau,2014Sridharan,2017Ching}, the interactions between gas dynamics and magnetic fields through various MHD processes such as ambipolar diffusion \citep{2001Ostriker,2008Li,2009Basu,2015CO}, reconnection diffusion \citep{2012Lazarian,2016XL,2017Kowal,2017Mocz}, magnetic braking \citep{2008ML,2011Li,2012Seifried}, 
and magnetic-field-rotation misalignment \citep{2010CH,2012Joos,2013Li} should also play important roles in the cores.

To study the interactions between gas dynamics and magnetic fields, we first investigate the alignment between velocity gradients and magnetic fields.
In strong MHD turbulence, the motion of Alfv\'{e}nic turbulent eddy is preferentially perpendicular to the direction of local magnetic fields \citep{1995GS,1999LV,2000CV}. 
Owing to this preferential direction of motion, MHD simulations suggest a velocity gradient technique to probe magnetic field orientations based on the expectation of perpendicular alignment between velocity gradients and magnetic fields in the regions without significant self-gravity \citep{2017GL,2018LY}.
The application of the velocity gradient technique to GALFA HI and PLANCK polarization data further reveal a perpendicular alignment between velocity gradients and magnetic fields in the diffuse regions of interstellar medium \citep{2017YL}.
For the regions dominated by self-gravity, collapsing material is expected to drag relatively weak magnetic fields, and the velocity gradients hence are suggested to become parallel aligned to magnetic fields \citep{2017YLb,2018LY}. In this work, we compare the velocity gradients obtained from molecular line data to the magnetic fields obtained from dust polarizations in the self-gravitating massive dense cores.

In addition, we study the linewidth difference between neutral and ionic molecular line profiles. 
In low ionization medium, ions are tied to magnetic fields, and neutrals are weakly decoupled from magnetic fields.
This decoupling, or ambipolar diffusion \citep{1956MS}, allows neutrals to move faster than ions, resulting in a wider linewidth of neutrals than that of ions \citep{1991Shu}.
The early observational works of \citet{2000aHoude,2000bHoude} and \citet{2003Lai} showed that the molecular ions indeed exhibit narrower line profiles compared to those of coexistent neutral molecules. 
\citet{2008LH} later established a new technique to estimate magnetic field strength through comparing the linewidths of ionic and neutral line profiles. 
They noted a constant difference between the HCN/HCO$^+$ 4--3 velocity dispersions in M17 at different spatial scales (i.e., velocity dispersion spectrum).
They interpreted the constant difference as the energy dissipation through ambipolar diffusion and estimated a plane-of-the-sky magnetic field strength of $\sim$ 1 mG in M17.
The analysis of velocity dispersion spectrum has been performed in the massive dense cores Cyg-N44 (alias DR21(OH)) and Cyg-N53 of the DR21 filament, revealing magnetic field strengths of 1.7 and 0.76 mG in the two cores, respectively \citep{2010Li,2010Hezareh,2014Hezareh}. The magnetic field strength of Cyg-N44 derived from the velocity dispersion spectrum is consistent with the 1.7--2.1 mG derived from the angular dispersion function of dust polarization observations \citep{2013Girart,2017Ching}.
In this work, we attempt to measure the magnetic field strengths of the second to the sixth most massive cores in the DR21 filament.

\begin{figure}
\includegraphics[scale=1]{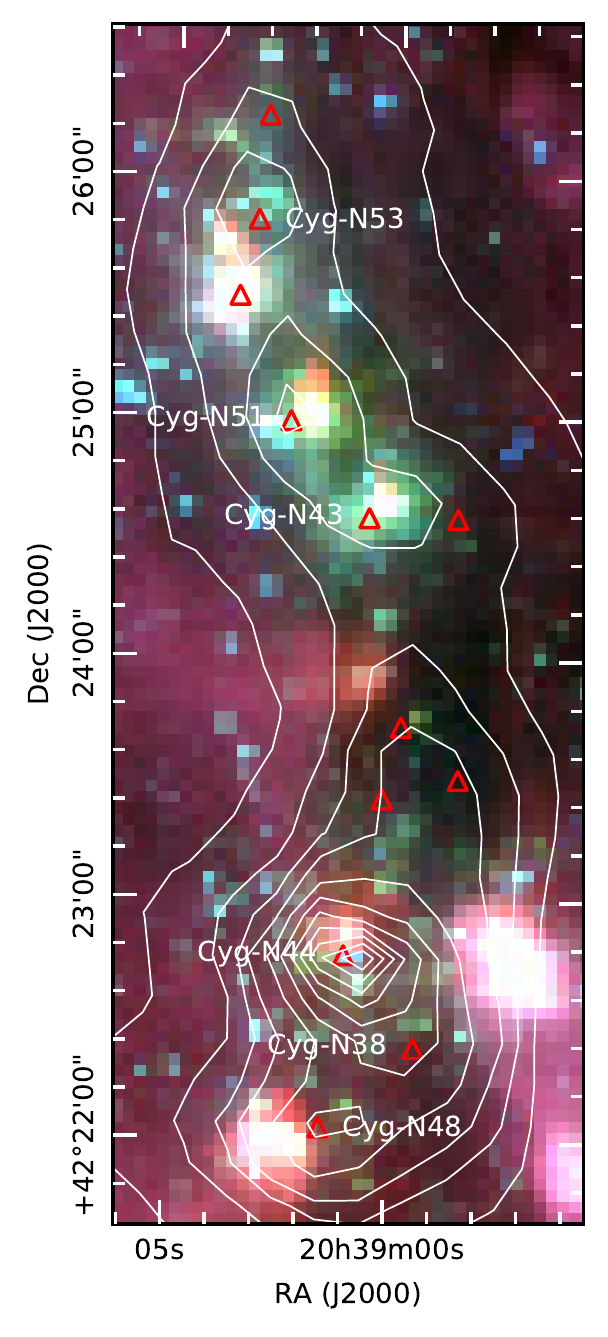}
\caption{A composite color image of DR21 filament of {\it Spitzer} 3.6 $\mu$m(blue), 4.5 $\mu$m(green) and 8.0 $\mu$m(red) images \citep{2006Smith}. Contours represent the 850 $\mu$m emission mapped using SCUBA on JCMT \citep{2009Matthews}. The triangles denote the positions of massive dense cores identified in the 1.2 mm continuum map of \citet{2007Motte}. The six massive cores in this work are labelled with their names.}
\label{fig_rgb}
\end{figure}

DR21 filament (Figure \ref{fig_rgb}) is a well studied high-mass star-forming region, located inside the Cygnus X molecular cloud complex \citep{2007Motte,2011Roy} at a distance of 1.4 kpc \citep{2012Rygl}. {\it Herschel} observations unveil that DR21 filament has a ridge of 4 pc length and 15000 M$_\sun$ mass, connected by several subfilaments with masses between 130 to 1400 M$_\sun$ \citep{2012Hennemann}. 
Molecular line observations reveal global infall motions of the DR21 filament, driven by continuous mass flows from the subfilaments onto the main filament \citep{2010Schneider}. Dust continuum maps of the DR21 filament discovered 12 massive dense cores \citep{2007Motte}. The massive dense cores drive active outflows \citep{2007Davis,2007Motte,2013Duarte,2014Duarte} and masers \citep{1983BE,2000Argon,2005Pestalozzi}, indicating recent high- to intermediate-mass star formation. 

We present the Submillimeter Array (SMA) 880 $\mu$m dust polarization observations of massive dense cores in the DR21 filament in \citet[][hereafter Paper I]{2017Ching}, which shows that the magnetic energy decreases from the filament to the cores and the ratio of kinematic energy to magnetic energy in massive dense cores is about 6, indicating that kinematics is more important than magnetic fields in the formation of massive dense cores. Here, we present SMA molecular line observations in the 345 GHz band toward massive dense cores Cyg-N38 (alias DR21(OH)-W), Cyg-N43 (W75S-FIR1), Cyg-N48 (DR21(OH)-S), Cyg-N51 (W75S-FIR2), and Cyg-N53. 
The observations and data reduction are described in Section 2. Section 3 presents the spectra and maps of the detected molecular lines. 
In Section 4, we combine the molecular line data and the dust polarization data in Paper I to explore the interactions between gas dynamics and magnetic fields in the cores. We discuss the dynamic properties and the magnetic fields of the cores in Section 5 and outline the main conclusions in Section 6.

\section{Observations and Data Reduction}\label{sec_obs}
The five massive dense cores have been observed from 2011 to 2015 with the 
SMA\footnote{The Submillimeter Array is a joint project between the Smithsonian Astrophysical Observatory and the Academia Sinica Institute of Astronomy and Astrophysics and is funded by the Smithsonian Institution and the Academia Sinica.} \citep{2004Ho} 
in the 345 GHz band.
The pointing centers of the sources are listed in Table \ref{spl_table}. 
The dates, array configurations, sources, and calibrators of the observations were listed in Table 1 of Paper I.
The observations of Cyg-N38, Cyg-N48 and Cyg-N53 were carried out with a single receiver, and the observations of Cyg-N43 and Cyg-N51 were carried out with dual receivers\footnote{The dual-receiver observations of Cyg-N44 and Cyg-N53 were used in Paper I to produce dust polarization data but not used in this paper to produce molecular line data.}. 
The single-receiver observations had a bandwidth of 4 GHz in each sideband, covering the 332--336 GHz frequencies in the lower sideband and the 344--348 GHz frequencies in the upper sideband.
The dual-receiver observations utilize the correlator to detect four polarization correlations (RR, LL, RL, LR) simultaneously, thus only cover the 334--336 GHz and 344--346 GHz  frequencies.
With the ASIC correlator, the velocity resolutions of the single- and dual-receiver observations were 0.7 km s$^{-1}$ and 1.4 km s$^{-1}$, respectively. 

The data were calibrated in the IDL MIR package for flux, bandpass, and time-dependent gain, and were exported to MIRIAD for imaging processes.
The detailed description of the data reduction were reported in Paper I. 
To generate the molecular line maps, the continuum emission in the visibility data was removed with the MIRIAD task UVLIN. Self-calibration using strong continuum emission was applied to the molecular line data.
Table \ref{obs_table} gives the rest frequencies, transitions, upper energy levels, and critical densities of the molecular lines detected in the observations.
The synthesized beams, position angles (P.A.), and rms noises of the molecular line images presented in this paper are summarized in Table \ref{map_table}.

\section{Results}\label{sec_result}
\begin{figure}
\includegraphics[scale=1.0]{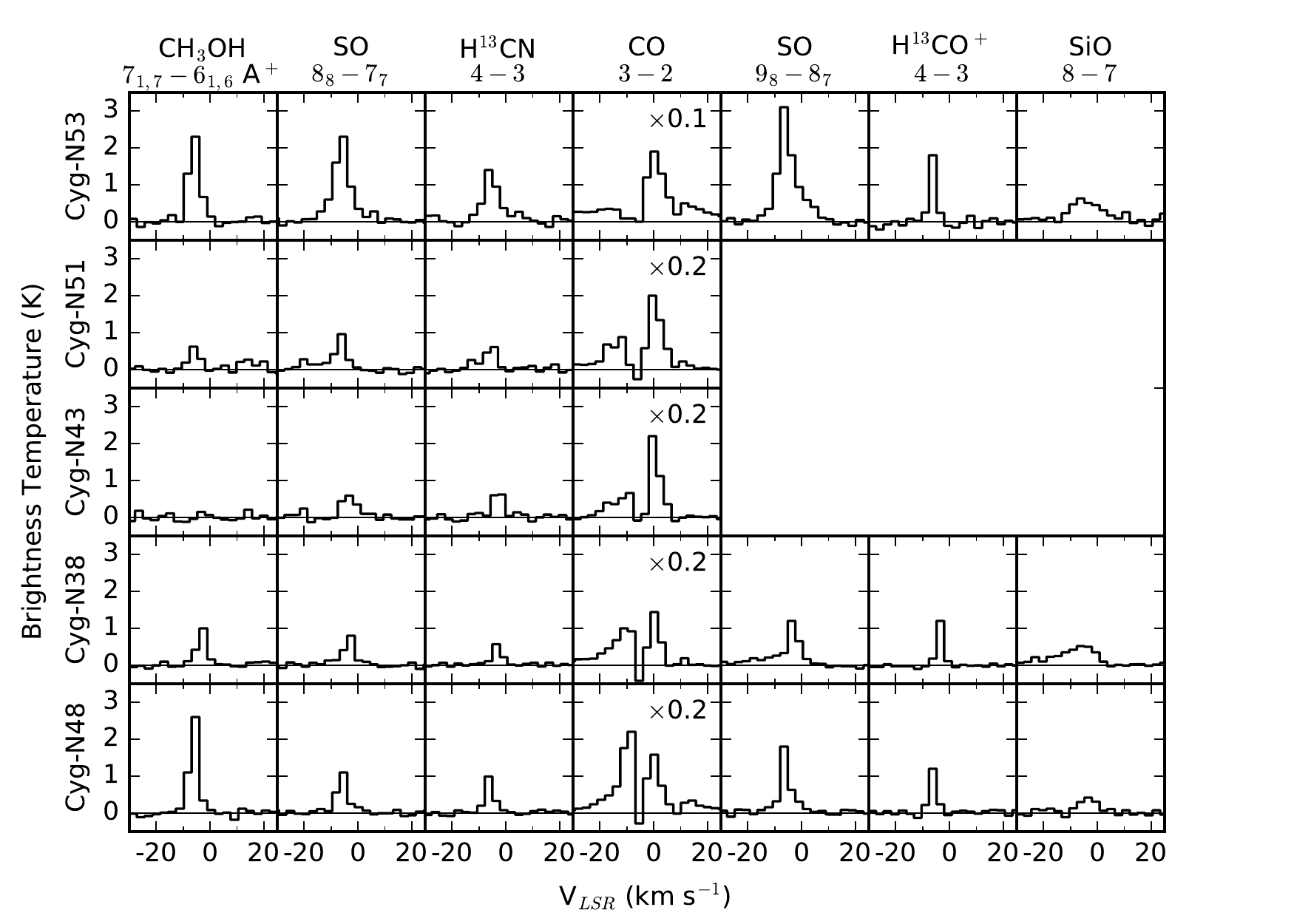}
\caption{Spectra of the detected lines in the five massive dense cores in the SMA 345 GHz band. The spectra are averaged over an area of $\sim$ 18 arcsec$^2$ around the continuum peaks of the cores. To compare the cores which were observed in different spatial and spectral resolutions, the spectra were uniformly resampled to a spatial resolution of 3.4$\arcsec$ and a spectral resolution of 2.8 km s$^{-1}$. The sources are plotted according to their delicnations, from north to south. The SO 9$_8$--8$_7$, H$^{13}$CO$^+$ 4--3, and SiO 8--7 lines were not covered in the bandwidth of the dual-receiver observations of Cyg-N43 and Cyg-N51.}
\label{fig_spec_grid}
\end{figure}

\begin{figure}
\includegraphics[scale=0.5]{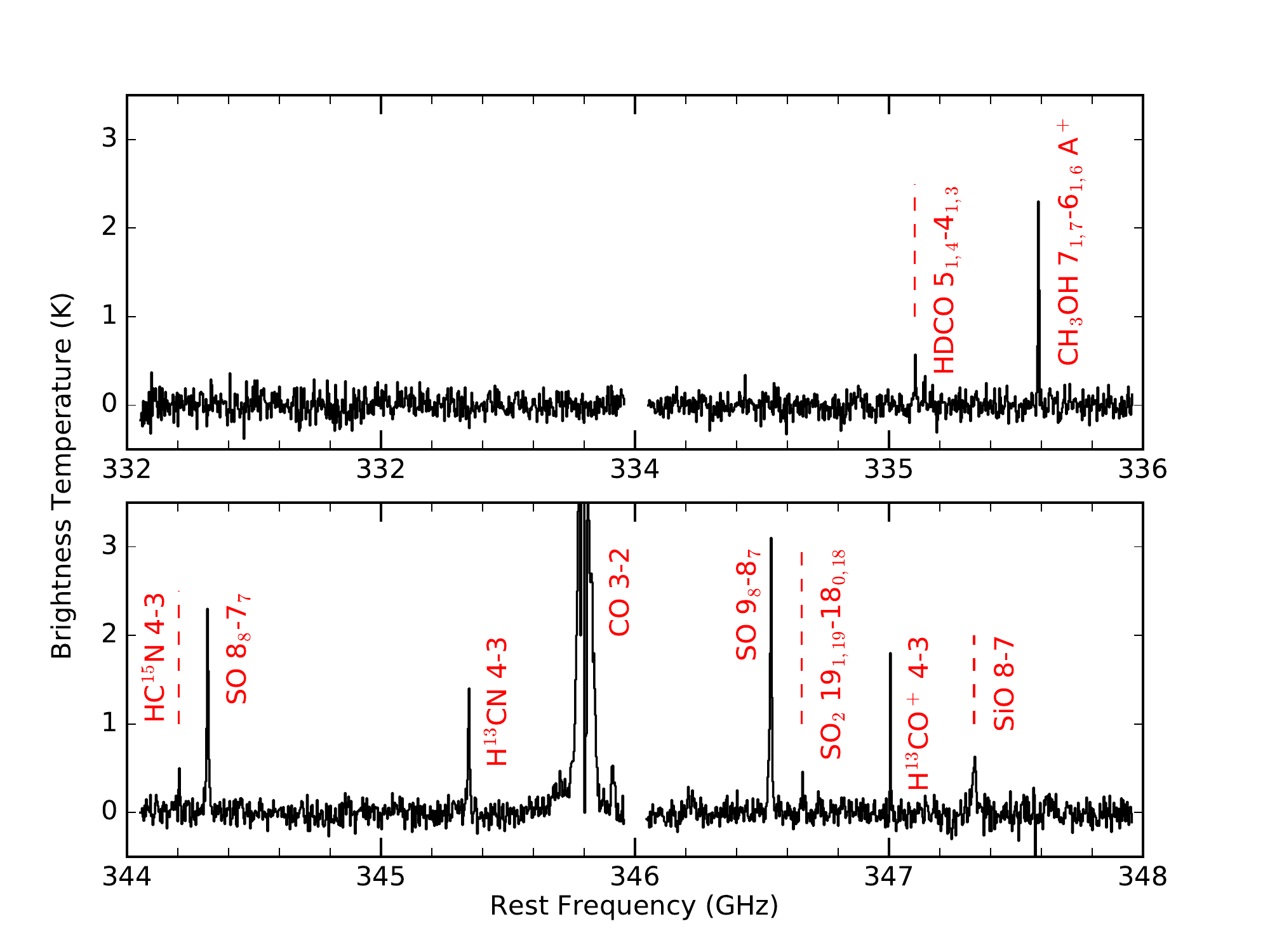}
\caption{Cyg-N53 SMA 345 GHz full bandwidth spectra averaged over an area of $\sim$ 18 arcsec$^2$ with a spatial resolution of 3.4$\arcsec$ and a spectral resolution of 2.8 km s$^{-1}$ around the continuum peak of the core.}
\label{fig_spec_N53}
\end{figure}

We present the spectra of the detected lines in the five massive dense cores in Figure \ref{fig_spec_grid}.
The molecular lines of CH$_3$OH 7$_{1,7}$--6$_{1,6}$ A$^+$, SO 8$_8$--7$_7$, H$^{13}$CN 4--3, and CO 3--2 are detected in all sources. SO 9$_8$--8$_7$, H$^{13}$CO$^+$ 4--3, and SiO 8--7 lines are detected in Cyg-N38, Cyg-N48, and Cyg-N53. 
Although the SO 9$_8$--8$_7$ line is blended with the SO$_2$ 16$_{4,12}$--16$_{3,13}$ line at 346.524 GHz (differ by $\sim$ 4 km s$^{-1}$), we denote that the detection should mostly come from SO molecule since the upper energy level of the SO$_2$ line is about two times higher than the SO line.
In spite of having a smaller bandwidth of the observations, the line emission of Cyg-N43 and Cyg-N51 is weaker than the other cores, indicating the poorest chemistry of the two cores. Cyg-N38 and Cyg-N48 appear to be slightly more rich than Cyg-N43 and Cyg-N51. Cyg-N53 is the chemically richest core among the five cores with the detections of HDCO 5$_{1,4}$--4$_{1,3}$, HC$^{15}$N 4--3,  and SO$_2$ 19$_{1,19}$--18$_{0,18}$ lines (Figure \ref{fig_spec_N53}), although Cyg-N53 is still less rich than Cyg-N44 in the DR21 filament \citep{2013Girart}.
The spectra of CH$_3$OH, SO, H$^{13}$CN, and H$^{13}$CO$^+$ molecules show narrow single-peaked profiles, whereas the CO and SiO molecules show broad profiles, indicating different origins of these two groups of molecules. 

\begin{figure}
\includegraphics[scale=0.8]{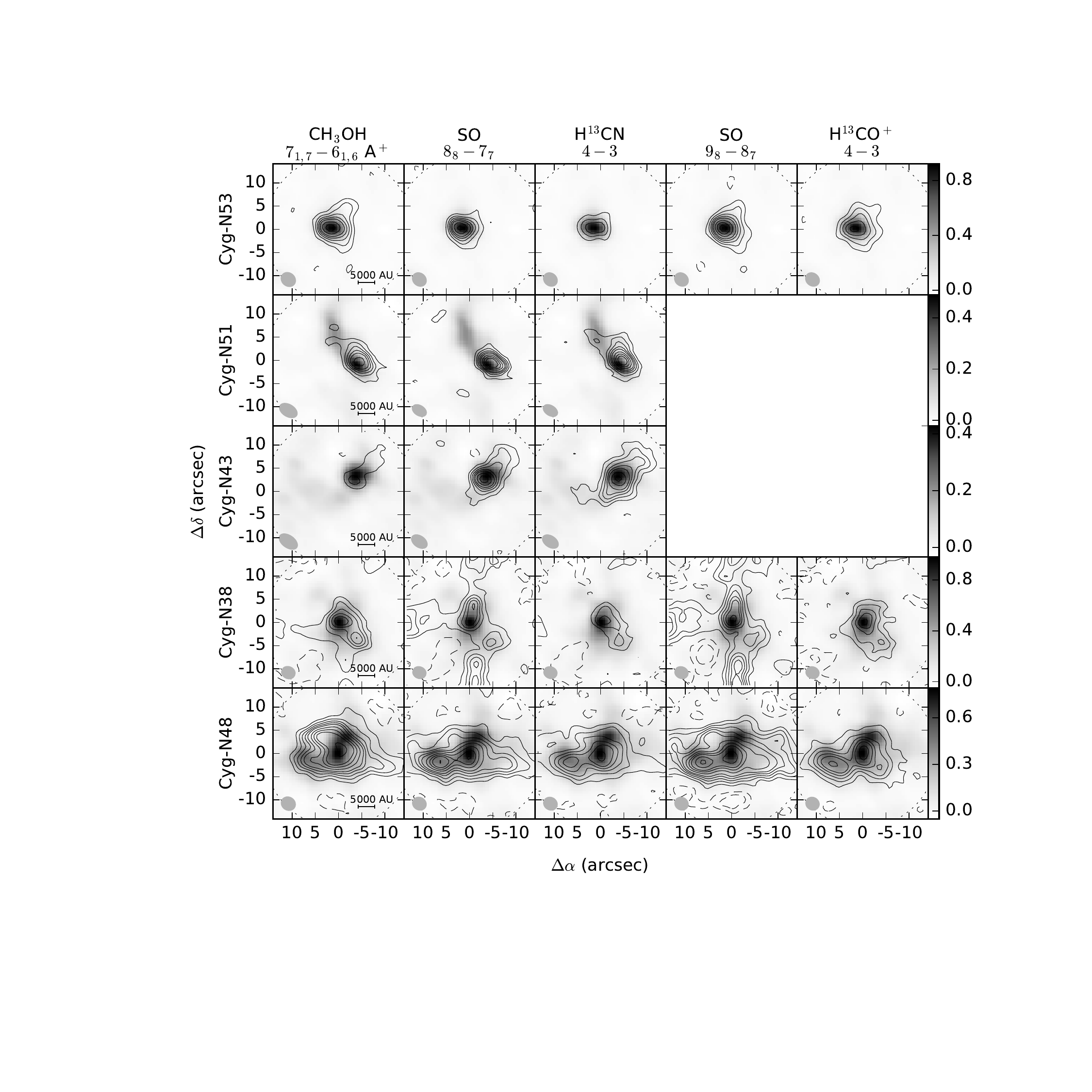}
\caption{Velocity integrated intensity (moment 0) maps of the CH$_3$OH 7$_{1,7}$--6$_{1,6}$ A$^+$, SO 8$_8$--7$_7$, H$^{13}$CN 4--3, SO 9$_8$--8$_7$, and H$^{13}$CO$^+$ 4--3 lines. The contour levels are -6, -3, 3, 6, 9, 12, 15, 20, 25, 30, 40, and 50 times the rms noises 
$\sigma_{\text{CH$_3$OH 7$_{1,7}$--6$_{1,6}$ A$^+$, N38}}$ = 0.36, $\sigma_{\text{SO 8$_8$--7$_7$, N38}}$ = 0.54, $\sigma_{\text{H$^{13}$CN 4--3, N38}}$ = 0.36, $\sigma_{\text{SO 9$_8$--8$_7$, N38}}$ = 0.53, $\sigma_{\text{H$^{13}$CO$^+$ 4--3, N38}}$ = 0.35, 
$\sigma_{\text{CH$_3$OH 7$_{1,7}$--6$_{1,6}$ A$^+$, N43}}$ = 0.32, $\sigma_{\text{SO 8$_8$--7$_7$, N43}}$ = 0.29, $\sigma_{\text{H$^{13}$CN 4--3, N43}}$ = 0.29, 
$\sigma_{\text{CH$_3$OH 7$_{1,7}$--6$_{1,6}$ A$^+$, N48}}$ = 0.38, $\sigma_{\text{SO 8$_8$--7$_7$, N48}}$ = 0.32, $\sigma_{\text{H$^{13}$CN 4--3, N48}}$ = 0.31, $\sigma_{\text{SO 9$_8$--8$_7$, N48}}$ = 0.35, $\sigma_{\text{H$^{13}$CO$^+$ 4--3, N48}}$ = 0.34,
$\sigma_{\text{CH$_3$OH 7$_{1,7}$--6$_{1,6}$ A$^+$, N51}}$ = 0.32, $\sigma_{\text{SO 8$_8$--7$_7$, N51}}$ = 0.35, $\sigma_{\text{H$^{13}$CN 4--3, N51}}$ = 0.26, 
$\sigma_{\text{CH$_3$OH 7$_{1,7}$--6$_{1,6}$ A$^+$, N53}}$ = 0.53, $\sigma_{\text{SO 8$_8$--7$_7$, N53}}$ = 0.58, $\sigma_{\text{H$^{13}$CN 4--3, N53}}$ = 0.63, $\sigma_{\text{SO 9$_8$--8$_7$, N53}}$ = 0.58, and $\sigma_{\text{H$^{13}$CO$^+$ 4--3, N53}}$ = 0.59 in units of Jy beam$^{-1}$ km s$^{-1}$.
The gray scale shows the SMA 880 $\mu$m continuum emission in units of Jy beam$^{-1}$.    
In each panel, the primary beam is shown as a dotted circle, and the synthesized beam is shown as a gray ellipse in the lower-left corner.}
\label{fig_mom0_core}
\end{figure}

Figure \ref{fig_mom0_core} shows the velocity integrated intensity maps of the CH$_3$OH, SO, H$^{13}$CN, and H$^{13}$CO$^+$ lines in the five massive dense cores, overlapped with the dust 880 $\mu$m continuum emission. 
The critical densities of the CH$_3$OH, SO, H$^{13}$CN, and H$^{13}$CO$^+$ lines are 10$^5$--10$^7$ cm$^{-3}$, which are similar to the average densities at an order of 10$^{6}$ cm$^{-3}$ derived from the dust emission (Paper I).
The emission of the \hcn and \hco lines exhibits compact morphology with sizes and peaks similar to those of dust emission, indicating that the \hcn and \hco lines trace dense cores better than the other lines.
Since the 80--90 K upper energy levels of the CH$_3$OH and SO lines are about a factor of 2 higher than those of the H$^{13}$CN and H$^{13}$CO$^+$ lines, the CH$_3$OH and SO lines trace warmer regions than the \hcn and \hco lines.
The CH$_3$OH and SO emission is more extended than the \hcn and \hco emission in Cyg-N38 and Cyg-N48, but the CH$_3$OH and SO emission is more compact than the \hcn and \hco emission in Cyg-N43 and Cyg-N51, suggesting that  Cyg-N38 and Cyg-N48 are warmer than Cyg-N43 and Cyg-N51.

Figure \ref{fig_mom0_CO} shows the velocity integrated intensity maps of the CO 3--2 and SiO 8--7 red- and blue-shifted emission in the massive dense cores. 
Owing to the complex structures in the low-velocity CO emission, the CO red- and blue-shifted intensity maps were integrated in a high-velocity range of $\sim$ 10--100 km s$^{-1}$ with respect to the system velocities of the sources, tracing narrow and elongated molecular outflows. 
On the contrary, SiO high-velocity emission is weak, and the SiO maps were obtained at velocities below $\sim$ 10 km s$^{-1}$ with respect to the system velocities.  
Below we present the dynamical features reveal by the molecular line data in the sources from north to south.

\subsection{Cyg-N53}
\begin{figure}
\includegraphics[scale=.9]{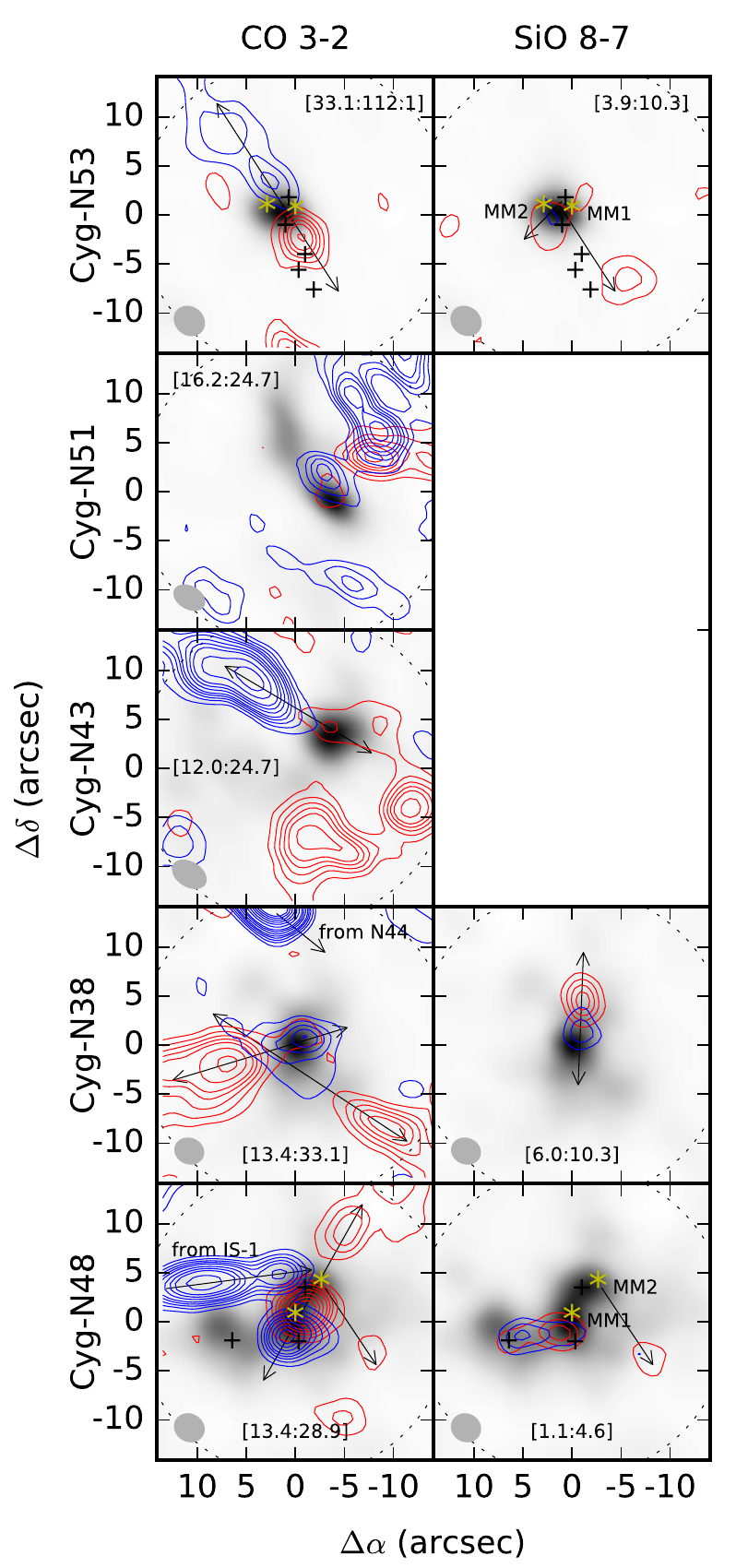}
\caption{Velocity integrated intensity maps of the red-shifted (red contours) and blue-shifted (blue contours) emission of the CO 3--2 (left panels) and SiO 8--7 (right panels) lines. The contour levels are 3, 6, 9, 12, 15, 20, 25, 30, 40, and 50 times the rms noises $\sigma_{\text{CO, N38}}$ = 0.76, $\sigma_{\text{SiO, N38}}$ = 0.31, $\sigma_{\text{CO, N43}}$ = 0.52, $\sigma_{\text{CO, N48}}$ = 0.70, $\sigma_{\text{SiO, N48}}$ = 0.29, $\sigma_{\text{CO, N51}}$ = 0.39, $\sigma_{\text{CO, N53}}$ = 4.2, and $\sigma_{\text{SiO, N53}}$ = 0.83 in units of Jy beam$^{-1}$ km s$^{-1}$. 
The velocity ranges used to generate the maps with respect to the systematic velocities of the sources are denoted in the brackets in units of km s$^{-1}$.
The gray scale shows the SMA 880 $\mu$m continuum emission.    
The arrows denote the axes of the identified outflows. The asterisks and crosses in Cyg-N48 and Cyg-N53 denote the major and minor fragments of the cores \citep{2010Bontemps}.}
\label{fig_mom0_CO}
\end{figure}

\begin{figure}
\includegraphics[scale=1.2]{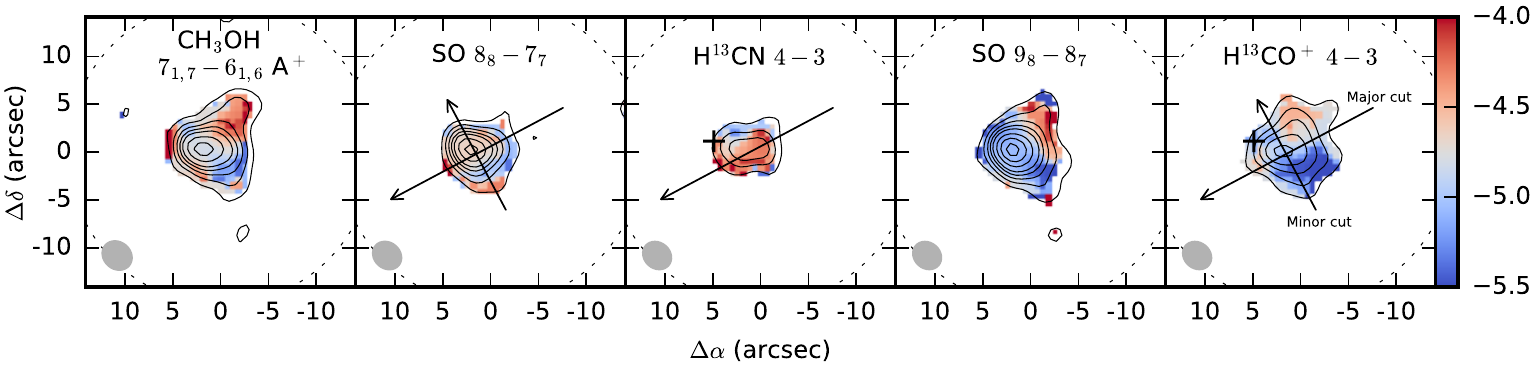}
\caption{Cyg-N53 CH$_3$OH 7$_{1,7}$--6$_{1,6}$ A$^+$, SO 8$_8$--7$_7$, \hcn 4--3, SO 9$_8$--8$_7$, and \hco 4--3 intensity-weighted velocity (moment 1) color images in units of km s$^{-1}$ overlapped with the velocity integrated intensity maps in contours. The arrows denote the PV cuts with the major cut at a P.A. of 118$^{\circ}$ determined by passing through the \hco 4--3 integrated intensity peak and being perpendicular to the NE-SW outflows. The crosses in the \hcn and \hco panels denote the position ($\alpha$, $\delta$)$_{\rm J2000}$ = (20$^h$39$^m$03\fs4, 42$^{\circ}$25$\arcmin$52$\arcsec$) with the minimum velocity dispersions.}
\label{fig_mom1_N53}
\end{figure}

\begin{figure}
\includegraphics[scale=1.2]{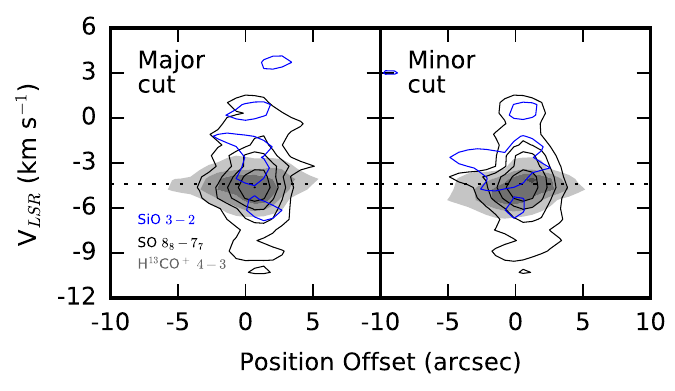}
\caption{ Cyg-N53 PV diagrams of the SO 8$_8$--7$_7$ (black), SiO 8--7 (blue), and \hco 4--3 (gray scale) lines. The contour levels are 3, 6, 9, 12, 15, and 20 times the rms noises in Table 3. The dotted line labels the systematic velocity of Cyg-N53.}
\label{fig_pv_N53}
\end{figure}

The CO 3--2 and SiO 8--7 maps (Figure \ref{fig_mom0_CO}) reveal a pair of red- and blue-shifted outflows in the northeast-southwest (NE-SW) direction. The direction and the morphology of the NE-SW outflows are consistent with those of the CO 2--1 and SiO 2--1 outflows originated from the fragment Cyg-N53 MM1 \citep{2013Duarte, 2014Duarte}.
Figure \ref{fig_mom1_N53} shows the intensity-weighted velocity maps in Cyg-N53.
There is no prominent velocity gradient in the core.
Figure \ref{fig_pv_N53} shows the PV diagrams of the dense core tracer \hco 4--3, warm gas tracer SO 8$_8$--7$_7$, and shock tracer SiO 8--7 perpendicular (major axis) and parallel (minor axis) to the NE-SW outflows. 
The PV diagrams reveal SO and SiO gas streams flowing between the core center and a +2--3$\arcsec$ position offset along the major axis, beyond the dense core revealed by the \hco emission. 
The maximum velocity of the streams is $\sim$ 8 km s$^{-1}$ with respect to the systematic velocity of Cyg-N53. The total mass of Cyg-N53 is 26 M$_\odot$ (Paper I), and the free-fall velocities at distances from 5000 to 1000 AU of a 26 M$_\odot$ central object are 3--7 km s$^{-1}$. 
Considering that the mass within 1000 AU of the core is much less than 26 M$_\odot$, the infall velocity of Cyg-N53 should be slower than 7 km s$^{-1}$.
Note that the spatial resolutions of the observations are 4000-4900 AU, and thus our data should not be sensitive to infalling gas within 1000 AU of the core owing to the limited resolution.
Since the SO and SiO gas streams are faster than the expected infall velocity along with a position offset from the line of sight of the core, the SO and SiO emission in Cyg-N53 is more likely tracing outflows in a direction perpendicular to the NE-SW outflows than tracing infalls, and we need observations with a better angular resolution to resolve the gas dynamics of the source. 
The SiO gas stream along the major axis appears to be the red contours at the southeast part of the core in our SiO 8--7 map (Figure \ref{fig_mom0_CO}) as well as in the SiO 2--1 map of \citet{2014Duarte}.

\subsection{Cyg-N51}
\begin{figure}
\includegraphics[scale=1.5]{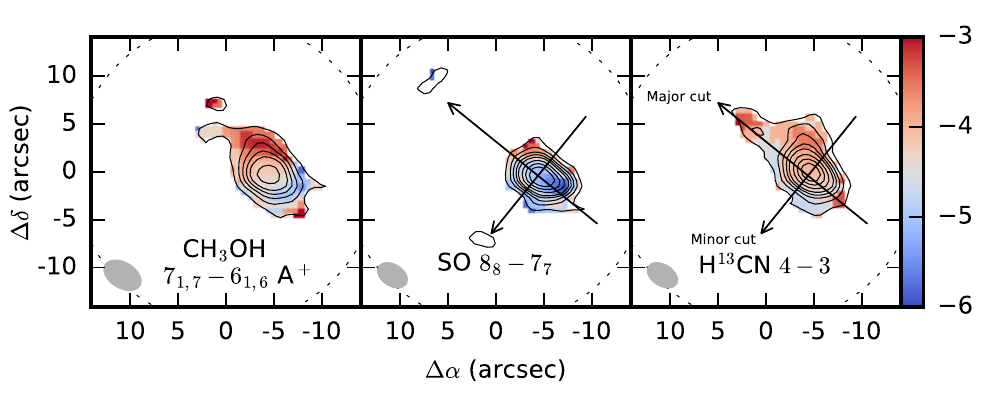}
\caption{Cyg-N51 CH$_3$OH 7$_{1,7}$--6$_{1,6}$ A$^+$, SO 8$_8$--7$_7$, and \hcn 4--3 intensity-weighted velocity (moment 1) color images in units of km s$^{-1}$ overlapped with the velocity integrated intensity maps in contours. The arrows denote the PV cuts with the major cut at a P.A. of 51$^{\circ}$ determined by passing through the H$^{13}$CN 4--3 integrated intensity peak and being perpendicular to the CO 3--2 outflows.}
\label{fig_mom1_h13cn_N51}
\end{figure}

\begin{figure}
\includegraphics[scale=1.5]{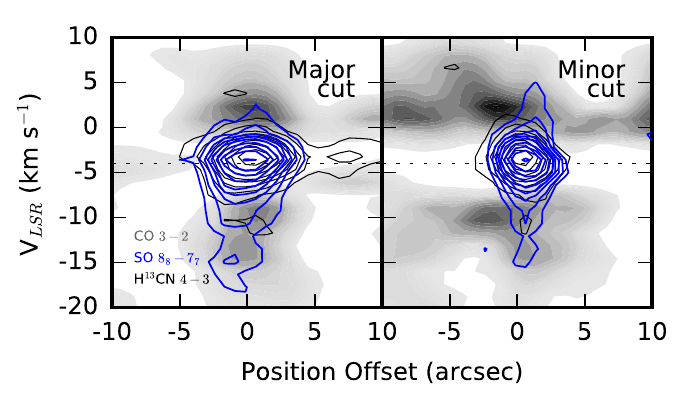}
\caption{ Cyg-N51 PV diagrams of the H$^{13}$CN 4--3 (black), SO 8$_8$--7$_7$ (blue), and CO 3--2 (gray scale) lines. The contour levels are 3, 6, 9, 12, 15, 20, 25, 30, and 40 times the rms noises in Table 3. The dotted line labels the systematic velocity of Cyg-N51.}
\label{fig_pv_N51}
\end{figure}

The CO red- and blue-shifted emission of Cyg-N51 (Figure \ref{fig_mom0_CO}) is weak at the center of the core and become strong at the northwestern side of the core, suggesting that the outflows might be launched in a direction close to the line of sight and be bent into the northwestern direction at a farther distance from the core. 
Figure \ref{fig_mom1_h13cn_N51} shows the intensity-weighted velocity maps of the source. 
There is no clear velocity gradient in the core. 
Figure \ref{fig_pv_N51} shows the PV diagrams perpendicular (major axis) and parallel (minor axis) to the outflow direction. 
While the \hcn PV diagrams appear to be structureless, the SO PV diagrams present flows with a maximum velocity of $\sim$ 13 km s$^{-1}$ with respect to the systematic velocity. This maximum velocity is larger than the 4--9 km s$^{-1}$ free-fall velocities at distances from 5000 to 1000 AU of Cyg-N51 with a total mass of 45 M$_\odot$ (Paper I). Meanwhile, the SO emission along the major axis appears to be have a position offset of +1--2$\arcsec$. The position offset of the SO emission is similar to the CO emission which traces outflows to farther distances and faster velocities. Therefore, we speculate that the SO emission of Cyg-N51 is more likely tracing outflows than infalls.

\subsection{Cyg-N43}
The CO 3--2 map of Cyg-N43 (Figure \ref{fig_mom0_CO}) reveals an asymmetric bipolar outflow in the NE-SW direction. The red-shifted lobe of the outflow exhibits a ``C'' shape morphology, which might be a result of a collision between the outflow and ambient gas or a contamination by other outflows. 

\begin{figure}
\includegraphics[scale=1.5]{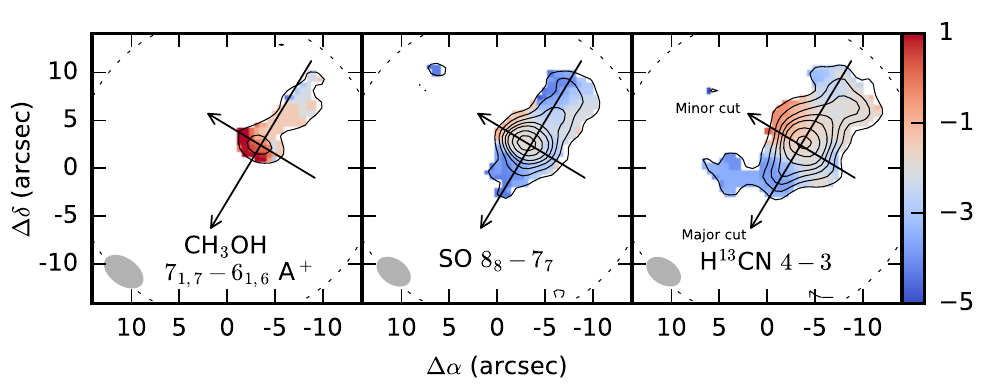}
\caption{Cyg-N43 CH$_3$OH 7$_{1,7}$--6$_{1,6}$ A$^+$, SO 8$_8$--7$_7$, and \hcn 4--3 intensity-weighted velocity (moment 1) color images in units of km s$^{-1}$ overlapped with the velocity integrated intensity maps in contours. The arrows denote the PV cuts with the major cut at a P.A. of 149$^{\circ}$ determined by passing through the H$^{13}$CN 4--3 integrated intensity peak and being perpendicular to the CO 3--2 outflows.}
\label{fig_mom1_h13cn_N43}
\end{figure}

\begin{figure}
\includegraphics[scale=1.5]{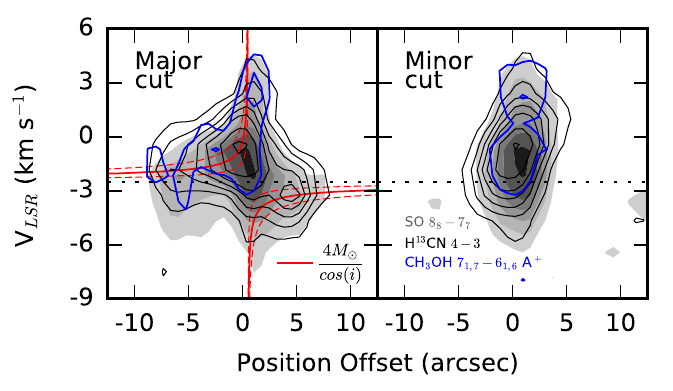}
\caption{ Cyg-N43 PV diagrams of the H$^{13}$CN 4--3 (black), CH$_3$OH 7$_{1,7}$--6$_{1,6}$ A$^+$ (blue), and SO 8$_8$--7$_7$ (gray scale) lines. The contour levels are 3, 6, 9, 12, 15, 20, and 25 times the rms noises in Table 3. 
The black dotted line labels the systematic velocity of Cyg-N43. The red lines show the PV curves of Keplerian rotation with a dynamical mass of 4 M$_\sun$/cos($i$). The red dashed lines represent the Keplerian rotation with masses ranging from 1 M$_\sun$/cos($i$) to 10 M$_\sun$/cos($i$).}
\label{fig_pv_N43}
\end{figure}

Figure \ref{fig_mom1_h13cn_N43} shows the intensity-weighted velocity maps in Cyg-N43. The \hcn intensity-weighted velocities present a velocity gradient perpendicular to the direction of the outflows. Meanwhile, the direction of the velocity gradient is almost parallel to the elongation of the H$^{13}$CN, SO, and dust emission. 
The velocity gradient is better revealed in the PV diagram along the major axis of the source (Figure \ref{fig_pv_N43}) that 
most of the red-shifted gas is distributed in the negative position offset, and most of the blue-shifted gas is distributed in the positive position offset. Meanwhile, the gas with a faster velocity distributes at a position closer to the core center. Both the features indicate a rotational motion of the core.

Since the red- and blue-shifted peaks in the PV diagrams are well resolved, we fit the data with assumptions of Keplerian rotations in order to study whether the motion of the core is rotation and to derive reasonable bound mass. In Figure \ref{fig_pv_N43}, we plot the expected PV curves along the major axis of a Keplerian disk with a dynamical mass of 4 M$_{\sun}$/cos($i$), where the 4 M$_{\sun}$ is the typical mass of the fragments inside the Cygnus-X massive cores \citep{2010Bontemps} and $i$ is the inclination angle of the rotation axis with respect to the plane of the sky. 
Since the dust and molecular line emission of Cyg-N43 appears relatively flattened, $i$ should be close to 0$^{\circ}$.
The red- and blue-shifted peaks and the bulges of the contours pass though the Keplerian curves, indicating rotational motions of Cyg-N43 with fragments of the core as the central objects of the rotation. However, the asymmetric emission in the PV diagram along the minor axis (Figure \ref{fig_pv_N43}) indicates that the rotation of Cyg-N43 is not perfectly Keplerian.
 The Keplerian-like rotation in a 0.1 pc size have been found in other massive cores IRDC 18223-3 \citep{2009Fallscheer} and Cyg-N44 \citep{2013Girart}. 
Meanwhile, the velocity gradient of 13 km s$^{-1}$ pc$^{-1}$ in Cyg-N43 is similar to the 18 km s$^{-1}$ pc$^{-1}$ in Cyg-N44 \citep{2013Girart}. However, the 0.1-pc Keplerian-like rotation may not represent the final rotation of the protostars, since the fragments inside these Keplerian-like structures usually have rotational axes not parallel to those of the 0.1-pc rotations. 

\subsection{Cyg-N38}
The CO and SiO maps (Figure \ref{fig_mom0_CO}) present multiple outflows of the source. The CO emission shows clear bipolar outflow in the northwest-southeast (NW-SE) direction with the peak of the blue-shifted outflow and the western part of the red-shifted outflow overlapped with the dust emission peak, suggesting that the NW-SE outflows may be launched from a protostar centered in the core. 
The SiO emission traces a more compact bipolar outflow in the north-south direction. The center of the bipolar outflow is located $\sim$ 4 arcsecs (5600 au) north of the core center.
There is a third possible bipolar outflow originated from a southern protostar of Cyg-N38, with strong CO red-shifted emission in the southwest (SW) part of the core and weak CO blue-shifted emission at the center of the core. 

\begin{figure}
\includegraphics[scale=1.2]{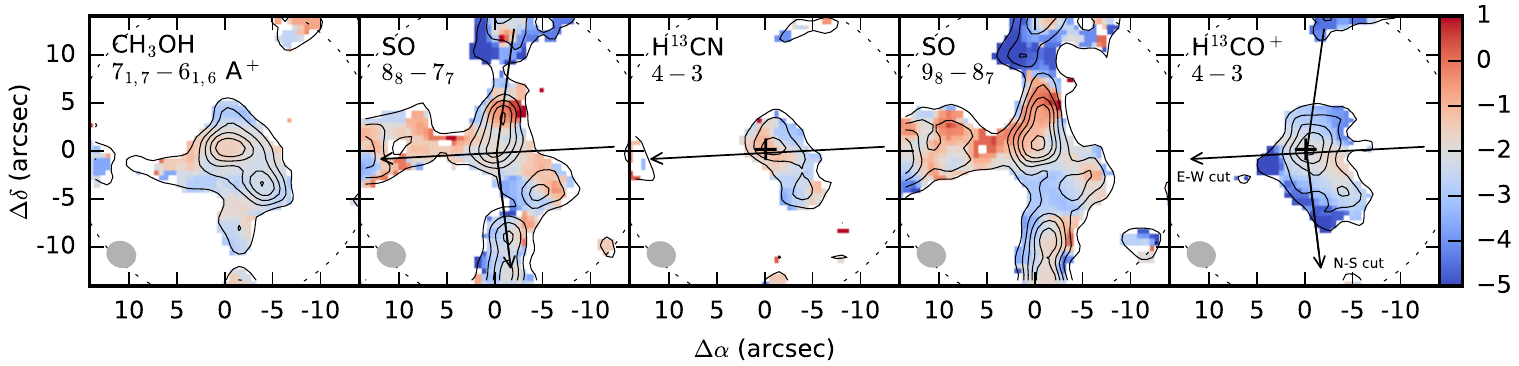}
\caption{Cyg-N38 CH$_3$OH 7$_{1,7}$--6$_{1,6}$ A$^+$, SO 8$_8$--7$_7$, \hcn 4--3, SO 9$_8$--8$_7$, and \hco 4--3 intensity-weighted velocity (moment 1) color images in units of km s$^{-1}$ overlapped with the velocity integrated intensity maps in contours. The arrows denote the PV cuts centered at the \hco integrated intensity peak with a P.A. of 93$^{\circ}$ for the E-W cut and a P.A. of 172$^{\circ}$ along the north filament and a P.A. of -173$^{\circ}$ along the south filament for the N-S cut. The crosses in the \hcn and \hco panels denote the position ($\alpha$, $\delta$)$_{\rm J2000}$ = (20$^h$38$^m$59\fs1, 42$^{\circ}$22$\arcmin$26$\arcsec$) with the minimum velocity dispersions.}
\label{fig_mom1_N38}
\end{figure}

\begin{figure}
\includegraphics[scale=1.5]{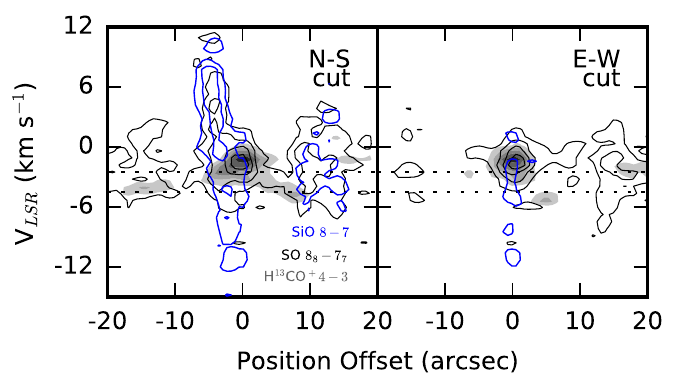}
\caption{ Cyg-N38 PV diagrams of the SO 8$_8$--7$_7$ (black), SiO 8--7 (blue), and 
 H$^{13}$CO$^+$ 4--3 (gray scale) lines. The contour levels are 3, 6, 9, 12, and 15 times the rms noises in Table 3.  The dotted lines label the range of -2.5 to -4.5 km s$^{-1}$ of the SO filamentary emission.}
\label{fig_pv_N38}
\end{figure}

Figure \ref{fig_mom1_N38} shows the intensity-weighted velocity maps in Cyg-N38. The dense core tracers \hcn and \hco exhibit no clear velocity gradient of the core. 
The SO and CH$_3$OH molecules present filamentary emission in the north, east, and south directions. 
Figure \ref{fig_pv_N38} shows the PV diagrams along the north-south (N-S) and the east-west (E-W) directions of the filamentary emission.
The distribution of the SO emission in the dense core is consistent with that of the \hco emission.
The high-velocity SO gas at $\sim$ 0$\arcsec$ position offset traces the SiO outflows, and the high-velocity SO gas at a +10--20$\arcsec$ position offset in the E-W cut traces the NW-SE CO outflows. Different to the structures of the dense core and the outflows, the SO emission along the filamentary structures reveal extended emission between a velocity range of $\sim$ -2.5 to -4.5 km s$^{-1}$.
The filamentary structures and the velocity ranges of our SO and CH$_3$OH results are similar to those of the convergent flows found in the N$_2$H$^+$ 1--0 emission \citep{2011aCsengeri}, but the velocity resolution of our data is not good enough to resolve the velocity components of the convergent flows. Considering that the SO and CH$_3$OH lines have upper energy levels higher than the other detected lines of Cyg-N38 presented in this work, the origin of the SO and CH$_3$OH filamentary emission could be associated with the colliding of the convergent flows.

\subsection{Cyg-N48}
In this source, the CO and SiO maps (Figure \ref{fig_mom0_CO}) also present multiple outflows.
The CO emission shows a bipolar outflow centered in Cyg-N48 MM1 with a P.A. of $-$28$^{\circ}$, which was already reported in the CO 2--1 maps in \citet{2013Duarte}. 
There is another red-shifted outflow in the SW direction, which is similar to the CO 2--1 and SiO 2--1 outflows launched form the fragment Cyg-N48 MM2 \citep{2013Duarte, 2014Duarte}.
The CO 3--2 blue-shifted emission in the northeastern part of Cyg-N48 is likely originated from the outflow of the Spitzer source IS-1 \citep{2014Duarte}.

In addition to the SW outflow, the SiO 8--7 emission (Figure \ref{fig_mom0_CO}) shows a strong component in the center of the core. The strong component is east-west elongated with the lowest contours of red- and blue-shifted emission overlapped with each other. Since the central region of the strong component do not overlap with any of the fragments of Cyg-N48, the strong component probably is not tracing outflows. 

\begin{figure}
\includegraphics[scale=1.2]{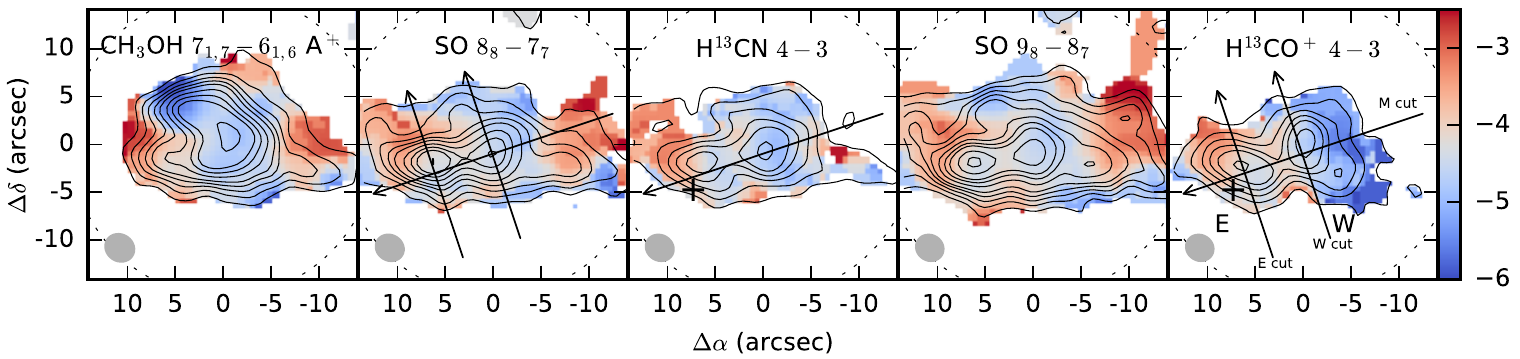}
\caption{Cyg-N48 CH$_3$OH 7$_{1,7}$--6$_{1,6}$ A$^+$, SO 8$_8$--7$_7$, \hcn 4--3, SO 9$_8$--8$_7$, and \hco 4--3  intensity-weighted velocity (moment 1) color images in units of km s$^{-1}$ overlapped with the velocity integrated intensity maps in contours. The two main peaks are labeled as E (east) and W (west). The arrows denote the PV cuts of the major axis (M cut), the east peak (E cut), and the west peak (W cut). The P.A. of the major axis is 109$^{\circ}$ determined by the elongation of the integrated \hco emission. The crosses in the \hcn and \hco panels denote the position ($\alpha$, $\delta$)$_{\rm J2000}$ = (20$^h$39$^m$02\fs2, 42$^{\circ}$22$\arcmin$00$\arcsec$) with the minimum velocity dispersions.}
\label{fig_mom1_h13cn_N48}
\end{figure}

\begin{figure}
\includegraphics[scale=1.5]{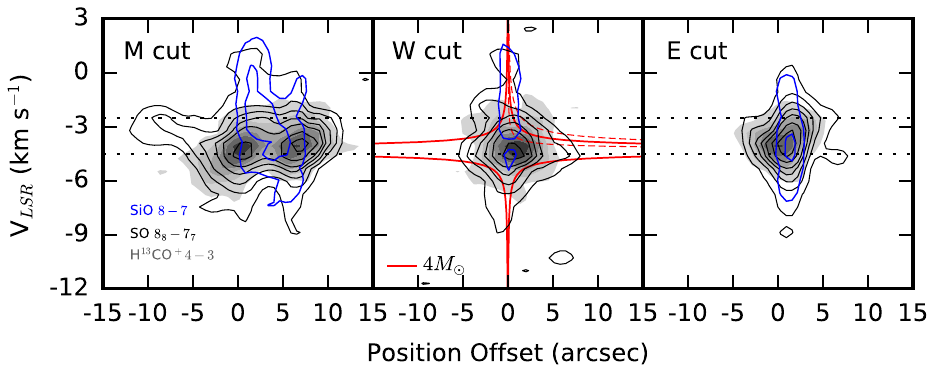}
\caption{ Cyg-N48 PV diagrams along the M, W, and E cuts of the SO 8$_8$--7$_7$ (black), SiO 8--7 (blue), and 
 H$^{13}$CO$^+$ 4--3 (gray scale) lines. The contour levels are 3, 6, 9, 12, 15, 20, 25, and 30 times the rms noises in Table 3. 
The red lines show the PV curves of free-fall velocities of a 4 M$_\sun$ central object. The red dashed lines represent the free-fall velocities with masses ranging from 1 to 10 M$_\sun$. The dotted lines label the range of -2.5 to -4.5 km s$^{-1}$ of the SO extended emission.}
\label{fig_pv_N48}
\end{figure}

Figure \ref{fig_mom1_h13cn_N48} shows the intensity-weighted velocity maps of the CH$_3$OH 7$_{1,7}$--6$_{1,6}$ A$^+$, SO 8$_8$--7$_7$, \hcn 4--3, SO 9$_8$--8$_7$, and \hco 4--3 lines. While the CH$_3$OH and SO molecules reveal complex velocity fields of Cyg-N48, the dense core tracers \hcn and \hco  exhibit a significant velocity gradient passing through the red-shifted eastern peak (labeled as E) and the blue-shifted western peak (labeled as W).
The \hco PV diagram along the major axis of Cyg-N48 presents a slope of 27 km s$^{-1}$ pc$^{-1}$ of the velocity gradient (Figure \ref{fig_pv_N48}). 
The E and W cores on the PV diagram also seem to have a similar slope, but the E and W cores are not aligned on the velocity gradient of the whole core. This indicates that the velocity fields of Cyg-N48 may have a combination of rotation and other gas dynamics (such as infalls and convergent flows, see below), instead of a simple pure rotating core. 

In Figure \ref{fig_pv_N48}, the SO 8$_8$--7$_7$ and SiO 8--7 emission along the line of sight of the E and W cores presents high-velocity components faster than the dense core tracer H$^{13}$CO$^+$.  
We plot the expected PV curves of free-fall velocities for a typical 4 M$_\sun$ fragment in the core, which appear to be similar to the elongated morphology of the SO and SiO emission at 0 position offset and at the systematic velocity of the source.  Nevertheless, a mass range of 1 to 10 M$_\sun$ can produce similar PV curves, and our data cannot tell whether the central source of this possible infall is the core as a whole (the total mass of Cyg-N48 is 102 M$_\sun$ within a FWHM size of 14000 AU, Paper I) or the fragments inside the core.
If the SO and SiO high-velocity emission is associated with outflows, the small position offset of the emission indicates that the outflow direction should be close to the line of sight. Since the previous studies of common outflow tracers CO and SiO with $\sim$ 1$\arcsec$ resolution have not found line-of-sight outflows in the E and W cores \citep{2013Duarte,2014Duarte}, the SO and SiO high-velocity emission is more likely associated with infalls than outflows.
The global infall of Cyg-N48 has been found in the \hco emission with a 28$\arcsec$ ($\sim$ 0.2 pc) resolution \citep{2011bCsengeri}, and our SO and SiO maps is probably revealing small-scale infalls inside the E and W cores.  
The SiO molecule can be produced by the destruction of dust grain in shocks \citep{1997Schilke,2001Le,2008Gusdorf}. While SiO emission is commonly found in outflows, the SiO emission associated with infalls has been found in \citet{2016Girart} and \citet{2017Juarez}.

The intensity maps of the SO and CH$_3$OH molecules present emission more extended than those of the dense core tracers \hcn and H$^{13}$CO$^+$ (Figure \ref{fig_mom0_core}). In Figure \ref{fig_pv_N48}, the morphology of the SO emission in the the high-density regions is consistent with the \hcn emission, and the SO emission in the extended regions have a velocity range of $\sim$ -2.5 to -4.5 km s$^{-1}$. 
Similar to the SO and CH$_3$OH emission in Cyg-N38, the extended structures and the velocity ranges of the SO and CH$_3$OH emission in Cyg-N48 are similar to those of the N$_2$H$^+$ 1--0 convergent flows found in \citet{2011aCsengeri}, suggesting that the SO and CH$_3$OH emission could be associated with the colliding of the convergent flows.

\section{Analysis}\label{sec_analysis}
\subsection{Alignment between Magnetic Fields and Velocity Gradients}

\begin{figure}
\includegraphics[scale=1]{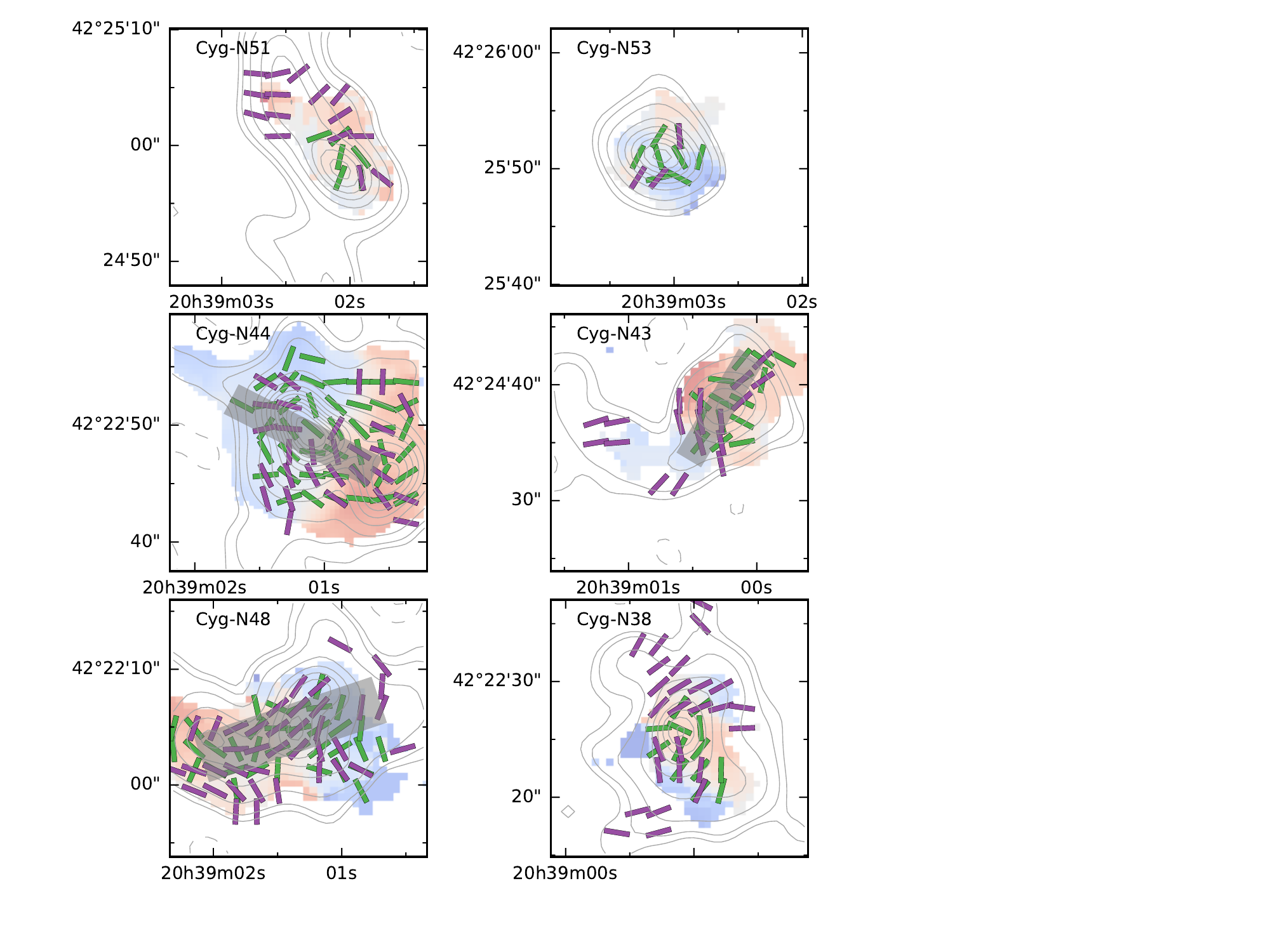}
\caption{Magnetic field orientations (magenta segments) and velocity gradient orientations (green segments) in the six massive dense cores of DR21 filament. 
The contours represent the 880 $\mu$m dust continuum emission. The color scales represent the intensity-weighted velocity maps of the sources. The shadowed regions represent the rotation-like regions of the cores.
Because the computation of one velocity gradient requires continuous detections of intensity-weighted velocities in seven pixels in the R.A. direction and in the Dec. direction, the green segments only appear over partial regions of the intensity-weighted velocity maps.}
\label{fig_mom1_pol_6maps}
\end{figure}

In paper I, we found that the kinematic energy is about a factor of 6 higher than the magnetic energy. 
Although the kinematic energy overwhelms the magnetic energy, the magnetic fields and velocity gradients could be perpendicular aligned in the regions with strong turbulence \citep{2017GL, 2017YL, 2018LY} or parallel aligned in the regions dominated by self-gravity \citep{2017YLb,2018LY}. 
Thus, here we compare the orientations of magnetic fields to the orientation of velocity gradients to study the interactions between the gas dynamics and magnetic fields.
Figure \ref{fig_mom1_pol_6maps} displays the magnetic field orientations inferred from our dust polarization observations (Paper I) and the velocity gradient orientations derived from the \hco intensity-weighted velocity maps of Cyg-N38, CygN48, Cyg-N53 and the \hcn intensity-weighted velocity maps of Cyg-N43 and Cyg-N51, given that the velocities of the \hcn and \hco lines should represent the velocity fields of the cores better than other lines.
To derive velocity gradients, we select intensity-weighted velocities on the intensity-weighted velocity maps with a Nyquist sampling as the polarization segments. Next, the gradients are computed using second order accurate central differences in the interior points and first order accurate one-sides differences at the boundaries \footnote{The \texttt{numpy.gradient} routine of python.}.
In the analysis, we also include the polarization data and the \hco data of Cyg-N44 in \citet{2013Girart}. Figure \ref{fig_mom1_pol_6maps} shows that the magnetic fields and velocity fields have both complex patterns in the cores. 
The distribution of the position angles of the magnetic field orientations in each core has a standard deviation of $\sim$30$^{\circ}$--40$^{\circ}$ (Paper I), and the position angles of the velocity gradient orientations have a similar level of dispersion.

\begin{figure}
\includegraphics[scale=1]{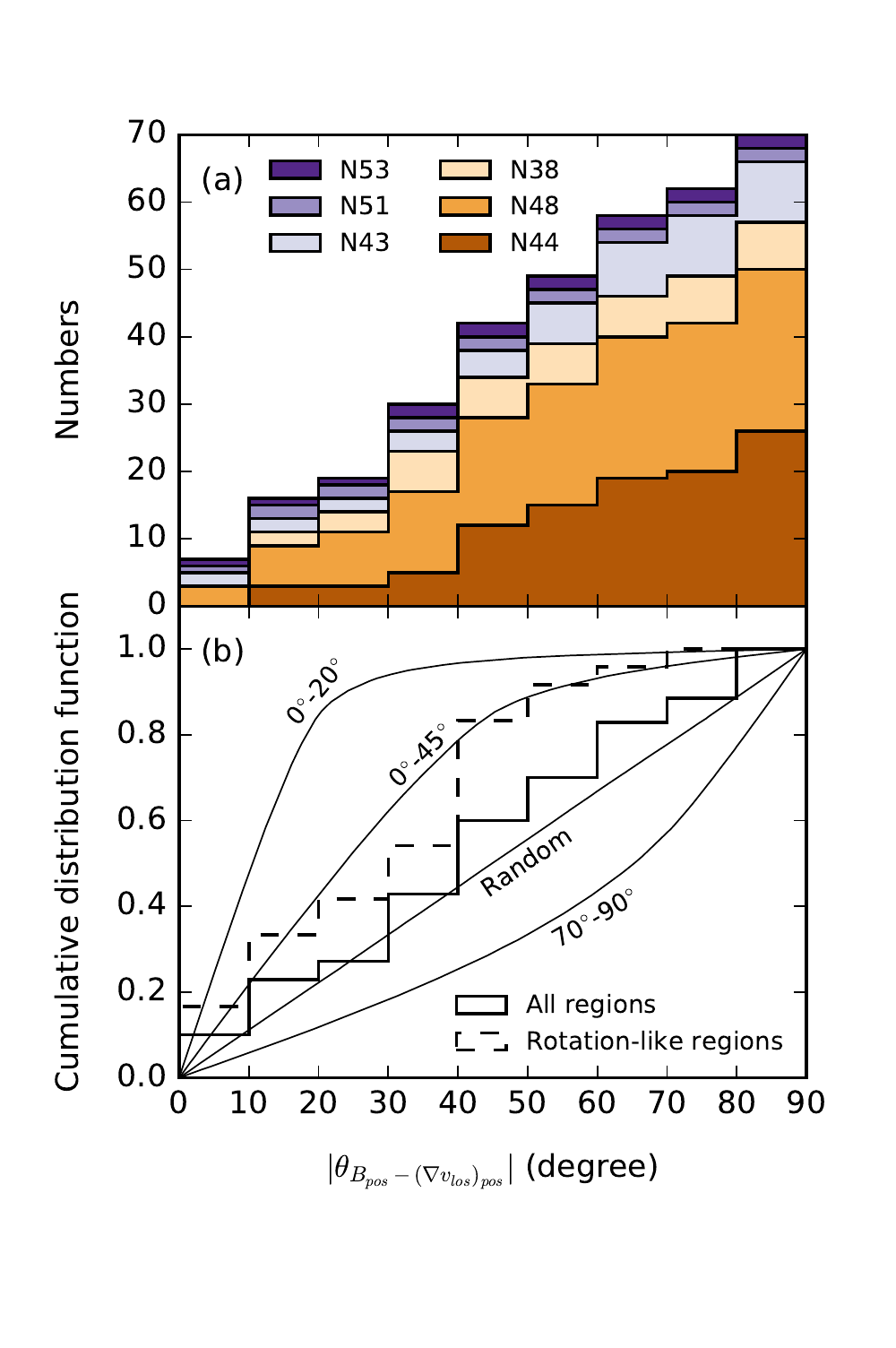}
\caption{(a) Cumulative histograms of $\theta_{B_{pos}-(\nabla v_{los})_{pos}}$ between the magnetic field orientations and the velocity gradient orientations in the cores. (b) Normalized cumulative histograms of $\theta_{B_{pos}-(\nabla v_{los})_{pos}}$. The solid and dashed stepped lines represent the histograms of all the $\theta_{B_{pos}-(\nabla v_{los})_{pos}}$ and the $\theta_{B_{pos}-(\nabla v_{los})_{pos}}$ in the shadowed regions in Figure \ref{fig_mom1_pol_6maps}, respectively. 
The curves are the cumulative distribution functions from Monte Carlo simulations of the magnetic field and gradient of $v_{los}$ aligned within 0$^\circ$--20$^\circ$, 0$^\circ$--45$^\circ$, and 70$^\circ$--90$^\circ$, or randomly aligned of one another.}
\label{fig_mom1_pol_hist}
\end{figure}

Figure \ref{fig_mom1_pol_hist} shows the cumulative histograms of the angular differences ($\theta_{B_{pos}-(\nabla v_{los})_{pos}}$) between the segments of magnetic fields and the segments of velocity gradients shown in Figure \ref{fig_mom1_pol_6maps}.
Note that the magnetic field revealed by dust polarization is the plane-of-the-sky component ($B_{pos}$), and the velocity gradient obtained from molecular line is the gradient of line-of-sight velocities ($v_{los}$).
Thus, $\theta_{B_{pos}-(\nabla v_{los})_{pos}}$ is the position angle between $B_{pos}$ and the plane-of-the-sky component of the gradient of $v_{los}$.
To study the projection effect from the magnetic field and gradient of $v_{los}$ in three-dimensional space to 
the $\theta_{B_{pos}-(\nabla v_{los})_{pos}}$ on the plane of sky, we adopt the Monte Carlo simulations of \citet{2014Hull}.
The simulations randomly select pairs of vectors in three dimensions that are aligned within 0$^{\circ}$--20$^{\circ}$, 0$^{\circ}$--45$^{\circ}$, 70$^{\circ}$--90$^{\circ}$, or random alignment of one another, and then measure the angular differences between the projected vectors on the plane of sky.
The resulting cumulative distribution functions of the simulations are shown in Figure \ref{fig_mom1_pol_hist}. 
The histogram of $\theta_{B_{pos}-(\nabla v_{los})_{pos}}$ in the six cores is more close to the simulations of random alignment than the other simulations, indicating that the magnetic fields and the gradients of $v_{los}$ of the massive cores do not strongly correlate to each other. 
We further study the alignment between magnetic fields and gradients of $v_{los}$ in the regions with significant velocity gradients. The shadowed regions in Figure \ref{fig_mom1_pol_6maps} represent the regions with maximum velocity gradients in Cyg-N43, Cyg-N44, and Cyg-N48, and the $\theta_{B_{pos}-(\nabla v_{los})_{pos}}$ obtained in the shadowed regions are plotted as the dashed histogram in Figure \ref{fig_mom1_pol_hist}. The dashed histogram is more close to the simulation of 0$^{\circ}$--45$^{\circ}$ alignment than the solid histogram, indicating that the magnetic fields and the gradients of $v_{los}$ are more likely be parallel in the regions with significant velocity gradient than the other regions of the cores. 

\begin{figure}
\includegraphics[scale=1]{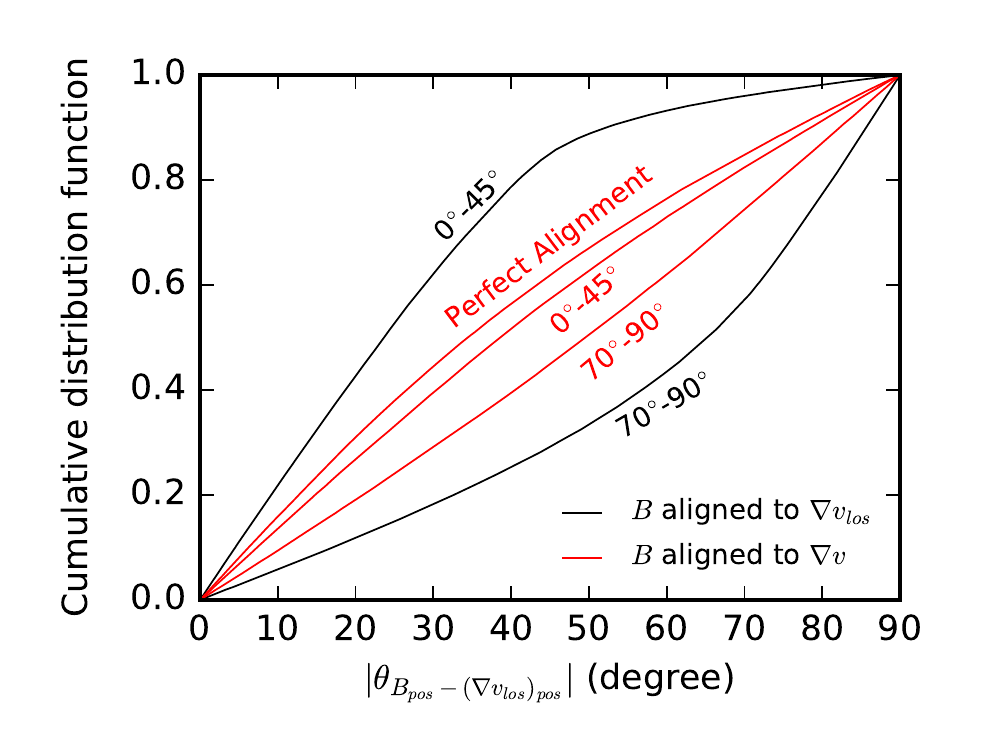}
\caption{Cumulative distribution functions from Monte Carlo simulations of the alignment between magnetic field and gradient of $v_{los}$ (black lines) and the alignment between magnetic field and gradient of total velocity (red lines).}
\label{fig_sim_Bpos_Vg}
\end{figure}

Note that the gradient of $v_{los}$ ($\partial v_z / \partial x$, $\partial v_z / \partial y$, $\partial v_z / \partial z$; where the $x, y, z$ represent the coordinates of R.A., Dec, and line of sight) is different from the gradient of total velocity ($\partial v_x / \partial x + \partial v_y / \partial x + \partial v_z / \partial x$, $\partial v_x / \partial y + \partial v_y / \partial y + \partial v_z / \partial y$, $\partial v_x / \partial z + \partial v_y / \partial z + \partial v_z / \partial z$),
and the simulations in Figure \ref{fig_mom1_pol_hist} study the alignment between magnetic fields and gradients of $v_{los}$, not the alignment between magnetic fields and gradients of total velocity. 
We carry out additional simulations to investigate whether the $\theta_{B_{pos}-(\nabla v_{los})_{pos}}$ can be used to study the alignment between magnetic fields and gradients of total velocity. First, we generate a three-dimensional vector composed by a set of nine random numbers to represent the $\partial v_x / \partial x$, $\partial v_y / \partial x$, $\partial v_z / \partial x$, $\partial v_x / \partial y$, $\partial v_y / \partial y$,  $\partial v_z / \partial y$, $\partial v_x / \partial z$, $\partial v_y / \partial z$, and $\partial v_z / \partial z$ in the gradient of total velocity. Second, we randomly generate another three-dimensional vector to represent the magnetic field. If the first vector and the second vector are aligned within 0$^{\circ}$--45$^{\circ}$ or 70$^{\circ}$--90$^{\circ}$, we measure the $\theta_{B_{pos}-(\nabla v_{los})_{pos}}$ between the vector ($\partial v_z / \partial x$, $\partial v_z / \partial y$) and $B_{pos}$. We also measure the $\theta_{B_{pos}-(\nabla v_{los})_{pos}}$ between the vector ($\partial v_z / \partial x$, $\partial v_z / \partial y$) and the vector ($\partial v_x / \partial x$ + $\partial v_y / \partial x$ + $\partial v_z / \partial x$, $\partial v_x / \partial y$ + $\partial v_y / \partial y$ +  $\partial v_z / \partial y$) to present a perfect alignment between magnetic field and gradient of total velocity.
The comparisons of the simulated $\theta_{B_{pos}-(\nabla v_{los})_{pos}}$ from the alignment of magnetic fields and gradients of $v_{los}$ and from the alignment of magnetic fields and gradients of total velocity are shown in Figure \ref{fig_sim_Bpos_Vg}. 
The difference between the 0$^{\circ}$--45$^{\circ}$ and 70$^{\circ}$--90$^{\circ}$ alignments in the simulations of magnetic fields and gradients of total velocity is less significant than those in the simulations of magnetic fields and gradients of $v_{los}$.
Meanwhile, for the same kind of alignment, the $\theta_{B_{pos}-(\nabla v_{los})_{pos}}$ from the simulations of magnetic fields and gradients of total velocity is always closer to the diagonal cumulative distribution function than the simulations of magnetic fields and gradients of $v_{los}$.
Since the diagonal cumulative distribution function represents random alignment, the $\theta_{B_{pos}-(\nabla v_{los})_{pos}}$ is not very sensitive to distinguish whether magnetic fields and gradients of total velocity are parallel aligned, randomly aligned, or perpendicular aligned.

The histogram of the observed $\theta_{B_{pos}-(\nabla v_{los})_{pos}}$ is slightly higher than the diagonal cumulative distribution function. 
Although this histogram indicates that the magnetic fields tend to be randomly aligned to the gradients of $v_{los}$, 
this histogram cannot constrain whether the magnetic fields and the gradients of total velocity are parallel aligned or randomly aligned,   since both cases would generate magnetic fields that appear to be randomly aligned to the gradients of $v_{los}$. 
The histogram of the observed $\theta_{B_{pos}-(\nabla v_{los})_{pos}}$ in the regions with significant velocity gradients is higher than the simulations of perfect alignment between magnetic fields and gradients of total velocity.
This indicates that in the regions with significant velocity gradients, the correlation between the gradients of $v_{los}$ and the gradients of total velocity is stronger than the assumption used in the simulations of Figure \ref{fig_sim_Bpos_Vg} that the three components in the gradient of $v_{los}$ ($\partial v_z / \partial x$, $\partial v_z / \partial y$, $\partial v_z / \partial z$) is independent to the other six components ($\partial v_x / \partial x$, $\partial v_y / \partial x$, $\partial v_x / \partial y$, $\partial v_y / \partial y$,  $\partial v_x / \partial z$, $\partial v_y / \partial z$) in the gradient of total velocity.
The strong correlation between the gradients of $v_{los}$ and the gradients of total velocity could be associated with the structures of the gas dynamics. For example, mass  flows of disks could have gradients of $v_{los}$ parallel to the gradients of total velocity, and the toroidal magnetic fields of disks could also be parallel to the gradients of total velocity. 
The velocity gradients in Cyg-N44 would be rotation-like motions \citep{2013Girart}, and the velocity gradients in Cyg-N43 and Cyg-N48 would be a associated with rotations.
Therefore, our results of the preferentially parallel alignment between magnetic fields and gradients of $v_{los}$ in the regions with significant velocity gradient seem to be consistent with the observations of toroidal magnetic fields in the rotating cores \citep{2013Qiu,2013Liu,2014Rao,2014Stephens}.

\subsection{Comparison of H$^{13}$CN/H$^{13}$CO$^+$ linewidths}
\subsubsection{Formalism of deriving magnetic field strength}
We adopt the analysis of velocity dispersion spectrum proposed by \citet{2008LH} to estimate the plane-of-the-sky magnetic field strength of the massive dense cores.
\citet{2008LH} proposed that at length scales larger than the ambipolar diffusion scale, the velocity dispersion spectra of coexistent ions and neutrals share the same power index owing to the flux freezing condition above the ambipolar diffusion scale, but the ion spectrum is downshifted uniformly relative to that of the neutrals owing to the occurrence of ambipolar diffusion at small scales. They suggested Kolmogorov-type power-law equations to fit the velocity dispersion spectra of ions ($\sigma_i$) and neutrals ($\sigma_n$) as functions of length scale $L$ by 
\begin{equation}
\begin{aligned}
& \sigma_i^2 \left(L\right)=bL^n+a \\
& \sigma_n^2 \left(L\right)=bL^n,
\end{aligned}
\end{equation}
where the fitting parameters $a$, $b$, and $n$ can be used to determine the ambipolar diffusion decoupling length-scale $L_{AD}$ and the neutral velocity dispersion $V_{AD}$ following: 
\begin{equation}
\begin{aligned}
& L_{AD}^n=\frac{-a}{b\left(1-0.37n\right)} \\
& V_{AD}^2=a+bL_{AD}^n.
\end{aligned}
\end{equation}
\citet{2008LH} also defined a new effective magnetic Reynolds number $R_m = 4\pi n_i \mu \nu_i L V/B^2$ with ion density $n_i$, mean ion-neutral collision reduced mass $\mu$, and collision rate $\nu_i$ at characteristic velocity scale $V$ and length scale $L$. The Reynolds number represents the relative importance of magnetic flux freezing to magnetic diffusivity with $R_m \gg 1$ corresponding to flux freezing condition and $R_m < 1$ when ambipolar diffusion becomes important. Since the parameters $L_{AD}$ and $V_{AD}$ represent the scales where $R_m \sim 1$, by substituting $n_i = n_n \chi_e$ with ionization fraction $\chi_e$ and neutral volume density $n_n$, $\nu_i = 1.5 \times 10^{-9}\ n_n\ s^{-1}$ \citep{1984Nakano}, one can obtain the plane-of-the-sky magnetic field strength $B_{pos}$ as 
\begin{multline}
B_{pos}=\left(\frac{L_{AD}}{0.5\ mpc}\right)^{1/2}\left(\frac{V_{AD}}{1\ km\ s^{-1}}\right)^{1/2} \\
\left(\frac{n_n}{10^6\ cm^{-3}}\right) \left(\frac{\chi_e}{10^{-7}}\right)^{1/2}\ \text{mG}.
\end{multline}

\subsubsection{Velocity Dispersion Spectra of H$^{13}$CN/H$^{13}$CO$^+$ data}
\begin{figure}
\includegraphics[scale=1]{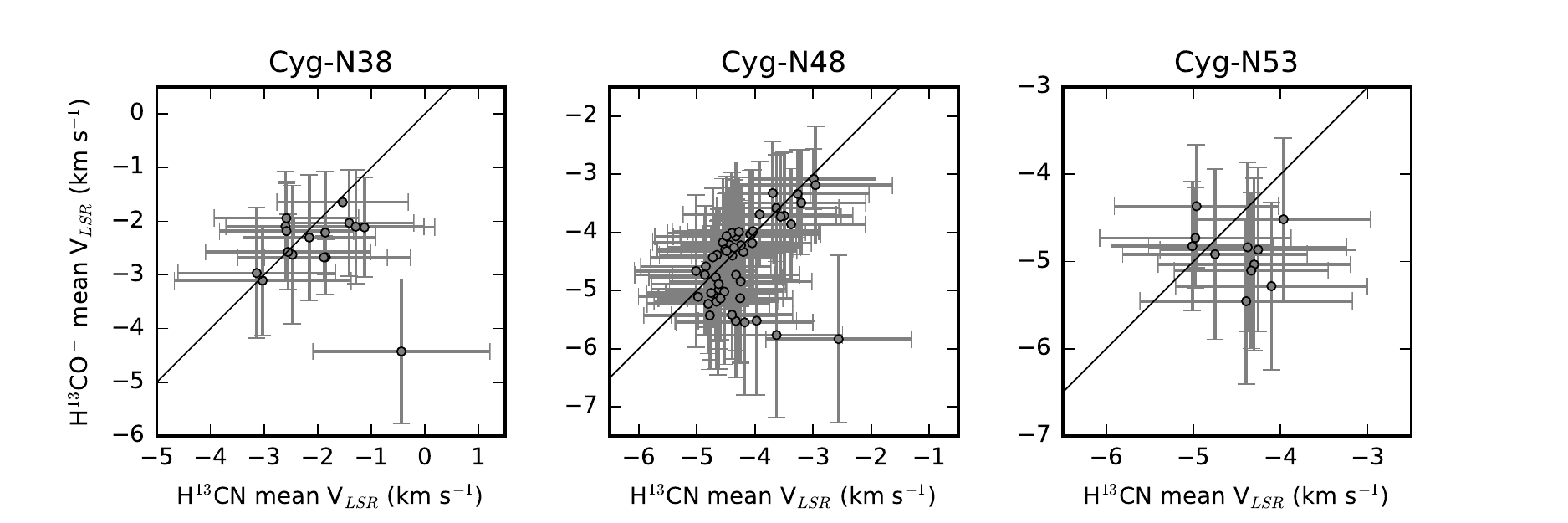}
\caption{Correlations between the intensity-weighted velocities of H$^{13}$CN and H$^{13}$CO$^+$ emission in Cyg-N38, Cyg-N48, and Cyg-N53. The linear correlation of $y= x$ is represented by a straight line in each panel.}
\label{fig_mom1_H13CX}
\end{figure}

\begin{figure}
\includegraphics[scale=0.8]{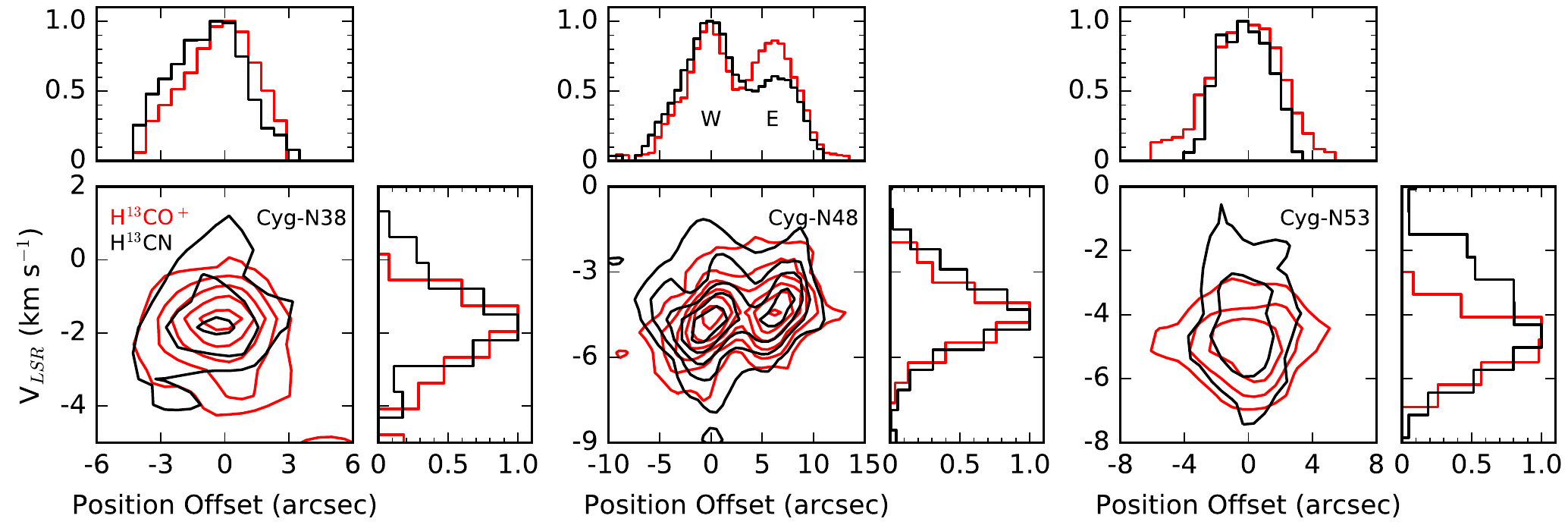}
\caption{\hcn 4--3 (black) and \hco 4--3 (red) PV diagrams along the E-W cut in Cyg-N38, the M cut in Cyg-N48, and the Major cut  in Cyg-N53. The contour levels are 3, 6, 9, 12, 15, and 20 times the rms noises in Table 3. For each source, the normalized spectra and the normalized velocity-integrated intensities of the two lines along the PV cuts are plotted at the right and top of the PV diagram, respectively.}
\label{fig_pv_H13CX}
\end{figure}

 We use our H$^{13}$CN and H$^{13}$CO$^+$ data to derive the velocity dispersion spectra and estimate magnetic field strengths of the sources. Figure \ref{fig_mom1_H13CX} shows the correlations between the intensity-weighted velocities of H$^{13}$CN and H$^{13}$CO$^+$ emission in Cyg-N38, Cyg-N48, and Cyg-N53. 
The H$^{13}$CN and H$^{13}$CO$^+$ emission generally follows a linear correlation in Cyg-N38 and Cyg-N48, but not in Cyg-N53.
Figure \ref{fig_pv_H13CX} shows the PV diagrams, normalized spectra, and normalized velocity-integrated intensities of the H$^{13}$CN and H$^{13}$CO$^+$ lines long the major PV cuts of the three sources. 
The PV diagrams of the two lines are consistent in Cyg-N38 and CygN48, but the \hcn line in Cyg-N53 traces velocities faster than those of the \hco line. 
The normalized spectra of the two lines have similar linewidths in Cyg-N38 and Cyg-N48, but again, the \hcn linewidth in Cyg-N48 is significantly wider than those of the \hco line owing to high-velocity  \hcn emission.
Except the E core of Cyg-N48, the normalized velocity-integrated intensities of the two lines show consistent profiles around the intensity peaks, suggesting that the two molecules are likely coexistent in the high-density regions of the cores despite that their critical densities differ by almost an order.
The coexistence of H$^{13}$CN/H$^{13}$CO$^+$ and their isotopes have also been suggested in the early works of comparing ionic and neutral line profiles \citep{2000aHoude,2000bHoude, 2002Houde,2003Lai} and been used in the studies of velocity dispersion spectrum \citep{2008LH,2010Li,2010Hezareh,2014Hezareh}. To obtain the velocity dispersion spectra at different scales, we smoothed the H$^{13}$CN and H$^{13}$CO$^+$ channel maps to spatial resolutions of 1.5, 2, and 2.5 times the synthesized beams. The velocity dispersions in the line profiles sampled at independent beams on the smoothed maps are obtained with gaussian fitting using nonlinear least-squares Marquardt-Levenberg algorithm for the H$^{13}$CO$^+$ 4--3 line and with the {\it HyperFine Structure} fitting routine in CLASS (or gaussian fitting if opacity $<$ 0.1) for the H$^{13}$CN 4--3 six hyperfine lines (Table \ref{table_h13cn}).
 
\begin{figure}
\centering
\includegraphics[scale=1.4]{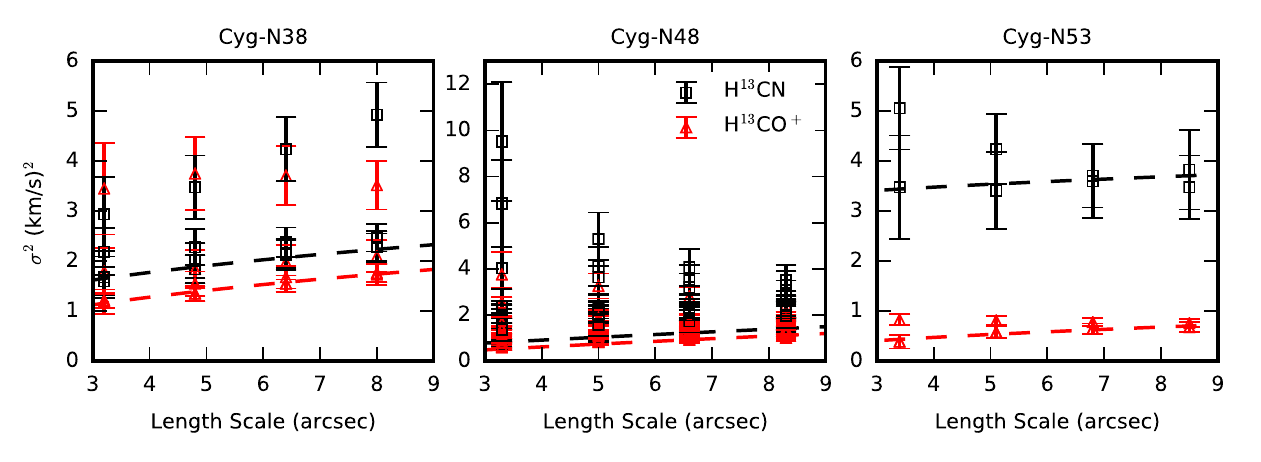}
\caption{Velocity dispersion spectra $\sigma^2$ of H$^{13}$CN (black) and H$^{13}$CO$^+$ (red) as a function of length scale $L$. The dashed lines represent the Kolmogorov-type laws $\sigma^2_n = bL^2$ to the H$^{13}$CN minimum $\sigma^2$ values and $\sigma^2_i = a + bL^2$ to the H$^{13}$CO$^+$ minimum $\sigma^2$ values.}
\label{fig_ambi_gauss}
\end{figure}

\begin{figure}
\includegraphics[scale=1.4]{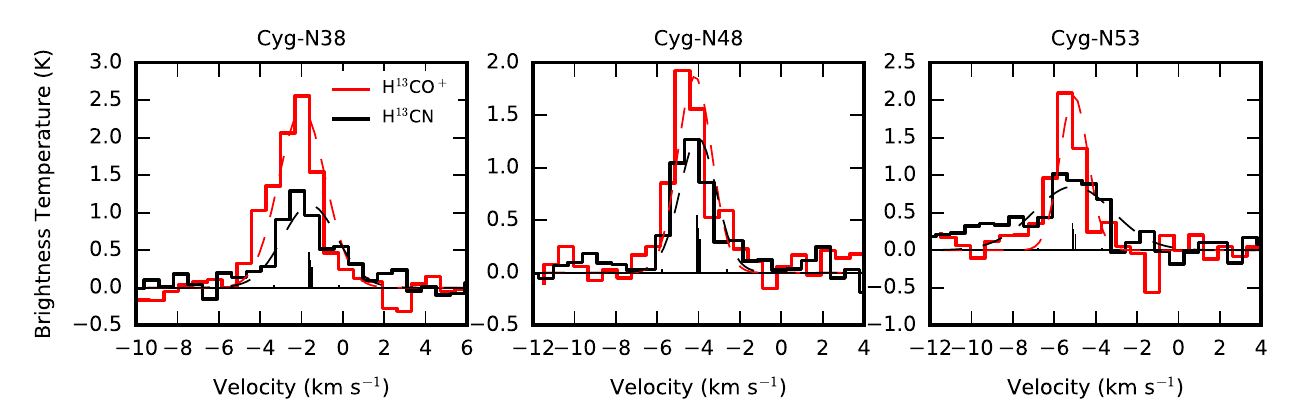}
\centering
\caption{
H$^{13}$CN (black) and H$^{13}$CO$^+$ (red) line profiles corresponding to the minimum $\sigma^2$ values at the smallest length scale. The dashed lines represent the fitted Gaussian profiles. The vertical segments denote the relative strengths of the H$^{13}$CN 4--3 six hyperfine lines.}
\label{fig_ambi_spec}
\end{figure}

Figure \ref{fig_ambi_gauss} shows the H$^{13}$CN/H$^{13}$CO$^+$ velocity dispersion spectra in Cyg-N38, Cyg-N48, and Cyg-N53. Here we plot the dispersions that are at least three times larger than the corresponding uncertainties and the difference between the central velocities of H$^{13}$CN and H$^{13}$CO$^+$ emission is smaller than the velocity resolution of our data (0.7 km s$^{-1}$).
Figure \ref{fig_ambi_spec} displays the H$^{13}$CN and H$^{13}$CO$^+$ line profiles at the positions with the minimum velocity dispersions $\sigma_n$ and $\sigma_i$ at the smallest length scale.
The positions with the minimum velocity dispersions are denoted as cross marks in Figure \ref{fig_mom1_N53}, \ref{fig_mom1_N38}, and \ref{fig_mom1_h13cn_N48}. In Cyg-N48, although the \hcn and \hco molecules appear to be not coexistent in the core E, the position with the minimum velocity dispersions locates in the low-density envelope, and hence the derived physical parameters using velocity dispersion spectrum should still be reliable.
In each source, the minimum values of the H$^{13}$CN and H$^{13}$CO$^+$ velocity dispersions show a power-law correlation to the length scales and a constant displacement between the H$^{13}$CN and H$^{13}$CO$^+$  spectra, in agreement with the predicted features in velocity dispersions owing to ambipolar diffusion in \citet{2008LH}. 

The Kolmogorov-type power-law fits to the spectra are represented as the dashed lines in Figure \ref{fig_ambi_gauss}, along with the fitting parameters ($a$, $b$, $n$) and the derived physical parameters ($V_{AD}$, $L_{AD}$, $B_{pos}$) listed in Table \ref{table_vds}. 
Here we adopt the volume density of the sources in Paper I as $n_n$ to derive $B_{pos}$ (Equation 3). 
The uncertainties in the $B_{pos}$ is about a factor of few, which is dominated by $\chi_e$ with a range of $\sim$ 10$^6$-10$^8$ \citep{1998Caselli}.
Cyg-N38 and Cyg-N48 have $L_{AD}$ $\sim$ 10 mpc and $B_{pos}$ about few mG, compatible to the results of 17--21 mpc and 0.7--1.7 mG derived from the velocity dispersion spectra in Cyg-N44 and Cyg-N53 \citep{2010Hezareh,2014Hezareh}.  
Interestingly, the parameters of $b = 0.241 \pm 0.026 $ and $n = 0.522 \pm 0.064$ of Cyg-N53 derived from the H$^{13}$CO$^+$ 1--0 data \citep{2014Hezareh} are in perfect agreement with our values using the H$^{13}$CO$^+$ 4--3 data, suggesting that the different H$^{13}$CO$^+$ transitions may have be emitted from an identical origin in Cyg-N53. However, the parameter $a$ of Cyg-N53 in our analysis seems to be affected by the bad correlation between H$^{13}$CN and H$^{13}$CO$^+$ emission (Figures \ref{fig_mom1_H13CX} and \ref{fig_pv_H13CX}) and the blue-shifted wing in the \hcn spectrum (Figure \ref{fig_ambi_gauss}). Thus $a$ of Cyg-N53 becomes an order of magnitude different to the $a$ in other cores of the DR21 filament, resulting in the extremely large values of $L_{AD}$ and $B_{pos}$ in Cyg-N53.
 
As compared to the 0.5--0.6 mG plane-of-the-sky magnetic field strengths derived from the dust polarization observations (Paper I), the analysis of velocity dispersion spectrum seems to always give stronger magnetic field strengths by at least a factor of 3. This might be a reasonable result for two reasons. First, the analysis of velocity dispersion spectrum assumes the ambipolar diffusion to be the only source of linewidth difference between ionic and neutral line profiles, but different spatial distributions or different excitation conditions of the ionic and neutral molecules may also cause different linewidths. For example, \hcn 4--3 line may trace regions with higher densities than \hco 4--3 line owing to the higher critical density of the former. If the gas dynamics such as outflows or infalls in the high-density regions are faster than those in the low-density regions, the linewidth of the \hcn line could become wider than that of the \hco line, resulting in an overestimation of magnetic field strength.   
Second, the strength derived from the dust polarization is an averaged strength of the core, but the strength derived from the velocity dispersion spectrum is a local value at the location of narrowest linewidth in both ionic and neutral line profiles. Therefore, we suggest that the magnetic field strength derived from velocity dispersion spectrum should be considered as an upper limit with respect to the averaged magnetic field strength in the core.

\section{Discussion}

In Paper I, we showed complex $B_{pos}$ in the massive dense cores, in contrast to the ordered $B_{pos}$ in the DR21 filament. We also derived $B_{pos}$ strength and estimated the kinematic, magnetic, and gravitational virial parameters. 
The $B_{pos}$ strength is 0.6 mG in the filament and 0.4--1.7 mG in the massive cores, and the magnetic energy per unit mass ($\mathcal{M}/M = \frac{1}{2}V_A^2$, where $V_A$ = $B$/$\sqrt{4\pi\rho}$ is the Alfv\'{e}n speed at density $\rho$) of the filament is about one order of magnitude higher than those of the cores.
The virial parameters show that the gravitational energy in the filament dominates magnetic and kinematic energies, whereas the kinematic energy in the cores dominates magnetic and gravitational energies.
The conclusions of Paper I that (1) the kinematics arising from gravitational collapse is more important than magnetic fields during the evolution from filaments to massive dense cores and (2) the magnetic energy is decreasing from the filament to the cores seem to be in good agreement with the analysis in Section 4.

First, the alignment between magnetic fields and velocity gradients in the massive cores is consistent with the finding that the kinematic energy is higher than the magnetic energy. 
Besides the possible rotation in Cyg-N43 and possible infall in Cyg-N43 revealed by our data, gas dynamics including outflows \citep{2013Duarte,2014Duarte}, infalls \citep{2011bCsengeri}, and convergent flows \citep{2011aCsengeri} are found in the massive cores of DR21 filament, which can be the sources of kinematic energy of the cores. 
Considering that there are usually more than one kind of gas dynamics in each of the core and the kinematic energy dominates the magnetic energy, the mix of the various kinds of gas dynamics would distort the ordered magnetic fields in the filament into the complex magnetic fields in the cores, making the magnetic fields and the velocity gradients appear to be randomly aligned in the cores. 
On the other hand, the mass flows in the regions with significant velocity gradients would represent rotational motions. If the kinematic energy is also larger than the magnetic energy in those regions, we speculate that the rotational motions would drag magnetic fields into toroidal structures, making the magnetic fields and the velocity gradients become parallel in regions with significant velocity gradients. Considering that the angular resolutions of this work were limited to resolve the rotational structures and the magnetic field structures inside the rotations, our hypothesis of toroidal fields in rotating cores should be tested with higher angular resolution observations. 

Second, the existence of ambipolar diffusion in small scales suggested by the analysis of velocity dispersion spectra is consistent with the finding of decreasing magnetic energy from the filament to the massive cores. 
The line profiles of ionic molecules are narrower than those of neutral molecules in the DR21 filament \citep{2000aHoude,2000bHoude,2003Lai}.
The velocity dispersion spectra of Cyg-N38, Cyg-N48, and Cyg-N53 in this work and those of Cyg-N44 and Cyg-N53 in \citet{2010Hezareh,2014Hezareh} are in agreement with the velocity dispersions owing to ambipolar diffusion predicted in \citet{2008LH}.
While the studies of ionic and neutral linewidths of the DR21 filament and the cores favor the ambipolar diffusion models, the highest spatial resolution ($\sim$ 0.02 pc or 5000 AU) of these studies still cannot resolve the length scale of ambipolar diffusion. This indicates that ambipolar diffusion might only exist in the high density regions ($>$ 10$^6$ cm$^{-3}$) of the cores. 
If ambipolar diffusion exists and only exists in the high density regions of cores, it should diffuse the magnetic energy of the cores into their parent filament, leading to a smaller magnetic energy of the cores than that of the filament. 

In MHD simulations, the alignment between magnetic fields and velocity gradients is expected to be perpendicular in the regions without significant self-gravity and be parallel in the regions dominated by self-gravity \citep{2017GL,2017YL,2017YLb,2018LY}. 
Our results of the random alignment between magnetic fields and gradients of $v_{lsr}$ suggest that magnetic fields and gradients of total velocity would be parallel aligned or randomly aligned in the massive dense cores.
Since the massive dense cores are significantly self-gravitating, the supercritical regions of the cores would have the collapsing material dragging magnetic fields and velocity gradients into parallel alignment, and the trans-critical regions of the cores would have the magnetic fields and velocity gradients appear to be randomly aligned owing to the transition from being perpendicular aligned to becoming parallel aligned. Therefore, our results seem to be consistent with the expectations of MHD simulations. In addition, our results rule out the perpendicular alignment between magnetic fields and velocity gradients, indicating that turbulence might not be the dominate mechanism regulating the directions of mass flows in the massive dense cores.

Our analysis of the alignment between magnetic fields and velocity gradients suggests that the gas dynamics would distort the magnetic field structures in the massive dense cores.
Meanwhile, the analysis of the velocity dispersion spectra suggests that ambipolar diffusion is likely important in dissipating the magnetic energies of the cores at scales smaller than 5000 AU. 
In high mass star-forming regions G35.2--0.74 N, G192.16--3.84, and  NGC 6334 V, the magnetic field lines are found to be likely dragged by gas dynamics such as rotations or convergent flows \citep{2013Liu, 2013Qiu, 2017Juarez}.
The existence of ambipolar diffusion in star-forming regions is also supported by the HCN and HCO$^+$ velocity dispersion spectra in M17 and NGC 2024 \citep{2008LH,2010Li}.
Therefore, the interactions between gas dynamics and magnetic fields through the process of the alignment between magnetic fields and velocity fields and the process of ambipolar diffusion would play important roles in determining the magnetic field structures and magnetic energies in massive dense cores.

\section{Conclusions}\label{sec_summary}
We present molecular line study of Submillimeter Array in the 345 GHz band of five massive dense cores in the DR21 filament.  
The main dynamical feature of the cores from north to south are the following:

\begin{enumerate}
\item  Cyg-N53 is the chemically richest core in our sources with the detections of HDCO 5$_{1,4}$--4$_{1,3}$, HC$^{15}$N 4--3,  and SO$_2$ 19$_{1,19}$--18$_{0,18}$ lines. A pair of red- and blue-shifted outflows in the northeast-southwest direction are revealed by the the CO and SiO maps, and a red-shifted outflow at the southeast part of the core is revealed by the SiO and SO maps. 

\item  The line emission of Cyg-N51 and Cyg-N43 is weaker than the other cores, indicating the poorest chemistry of the two cores. In Cyg-N51, outflows are revealed in the CO, SiO, and SO maps. In Cyg-N43, the CO map reveals outflows, and the CH$_3$OH, SO, and \hcn maps reveal a significant velocity gradient of the core. The velocity gradient of Cyg-N43 likely indicates a 0.1-pc-scale Keplerian-like rotation.

\item  The outflows of Cyg-N38 and Cyg-N48 are found in the CO and SiO maps. In Cyg-N48, possible infalls are found in the the SiO and SO maps, and a significant velocity gradient of the core are revealed by the \hcn and \hco lines.
Cyg-N38 and Cyg-N48 both show extended CH$_3$OH and SO filamentary emission between a velocity range of $\sim$ -2.5 to -4.5 km s$^{-1}$. Since the upper energy levels of the CH$_3$OH and SO lines indicate that the filamentary gas is warmer than the other regions, the filamentary emission seems to be associated with the N$_2$H$^+$ convergent flows previously found in the cores.
\end{enumerate} 

The velocity gradients of the cores derived from H$^{13}$CN and H$^{13}$CO$^+$ emission appear to be randomly aligned to the magnetic fields inferred from dust polarization, implying that the gradients of total velocity and magnetic fields are parallel aligned or randomly aligned in the three dimensional space.
In addition, there is a hint of 0$^\circ$--45$^\circ$ alignment between velocity gradients and magnetic fields in regions with significant velocity gradients, indicating that the rotational motions might lead to a stronger alignment between mass flows and magnetic fields than the other regions of the cores. 
Together with the finding in Paper I that the kinematic energy is higher the magnetic energy, the alignment between magnetic fields and velocity gradients suggests that the gas dynamics may distort the ordered magnetic fields in the filament into the complex magnetic fields in the cores, and observations with higher angular resolutions are needed to answer whether gas dynamics can determine directions of magnetic fields in the cores.

We perform an analysis of H$^{13}$CN/H$^{13}$CO$^+$ velocity dispersion spectra and derive the plane-of-the-sky magnetic field strengths of 1.9 mG and 5.1 mG in Cyg-N38 and Cyg-N48, respectively.
We suggest that the magnetic field strength derived from velocity dispersion spectrum should be considered as an upper limit with respect to the average magnetic field strength derived from dust polarization.
The existence of ambipolar diffusion suggested by the analysis of velocity dispersion spectra is consistent with the finding in Paper I of decreasing magnetic energy from the filament to the massive cores, suggesting that ambipolar diffusion is important in dissipating the magnetic energies of the cores at scales smaller than 5000 AU. 

\acknowledgments
Part of the data were obtained in the context of the SMA legacy project: Filaments, Magnetic Fields, and Massive Star Formation (PI: Qizhou Zhang).
T. C. C. acknowledges the support of the Smithsonian Predoctoral Fellowship, the University Consortium of ALMA-Taiwan (UCAT) Graduate Fellowship, and the Center for Astronomical Mega-Science, Chinese Academy of Sciences (CAMS).
T. C. C. and S. P. L. are thankful for the support of the Ministry of Science and Technology (MoST) of Taiwan through Grants 
102-2119-M-007-004-MY3, 104-2119-M-007-021, 105-2119-M-007 -022 -MY3, and 105-2119-M-007 -024.
Q. Z. acknowledges the support of the Scholarly Studies Awards ``Are Magnetic Fields Dynamically Important in Massive Star Formation?".
J. M. G. is supported by the MINECO (Spain) AYA2014-57369-C3 and AYA2017- 84390-C2 grants.
K. Q. acknowledges the support from National Natural Science Foundation of China (NSFC) through grants NSFC 11473011 and NSFC 11629302.
We also thank C. Hull for providing Monte Carlo simulations to study the alignment between magnetic fields and velocity gradients.

\clearpage	


\begin{deluxetable}{ccccc}
\tabletypesize{\scriptsize}
\tablecaption{Sample of Sources\label{spl_table}}
\tablewidth{0pt}
\tablehead{
\multirow{2}{*}{Source} & \multicolumn{2}{c}{Pointing Center} & $V_{LSR}$ & \multirow{2}{*}{Other name} \\
\cline{2-3} & $\alpha$ (J2000) & $\delta$ (J2000) & (km s$^{-1}$) & }
\startdata
Cyg-N38 & 20:38:59.11 & 42:22:25.96 & -2.5$^a$ & DR21(OH)-W \\
Cyg-N43 & 20:39:00.60 & 42:24:34.99 & -2.5$^a$ & W75S-FIR1 \\
Cyg-N44 & 20:39:01.20 & 42:22:48.50 & -3.0$^b$ & DR21(OH) \\
Cyg-N48 & 20:39:01.34 & 42:22:04.89 & -3.5$^c$ & DR21(OH)-S \\
Cyg-N51 & 20:39:02.40 & 42:24:59.00 & -4.0$^a$ & W75S-FIR2 \\
Cyg-N53 & 20:39:02.96 & 42:25:50.99 & -4.4$^c$ & -- 
\enddata
\tablenotetext{a}{Obtained from a Gaussian fit to the averaged H$^{13}$CN 4--3 spectrum of the entire core in this work.}
\tablenotetext{b}{\citet{2013Girart}} 
\tablenotetext{c}{\citet{2014Duarte}} 
\end{deluxetable}

\begin{deluxetable}{clllc}
\tabletypesize{\scriptsize}
\tablecaption{Detected Molecular Lines\label{obs_table}}
\tablewidth{0pt}
\tablehead{
Rest freq. & \multirow{2}{*}{Molecule} & \multirow{2}{*}{Transition} & $E_U$ & $n_{cr}^a$ \\
(GHz) & & & (K) & (cm$^{-3}$)} 
\startdata
335.097 & HDCO & 5$_{1,4}$--4$_{1,3}$ & 56 & -- \\
335.582 & CH$_3$OH & 7$_{1,7}$--6$_{1,6}$ A$^+$ & 79 & 3.2$\times$10$^5$ \\
344.200 & HC$^{15}$N & 4--3 & 41 & 1.5$\times$10$^7$ \\
344.311 & SO & 8$_8$--7$_7$ & 87 & -- \\
345.340 & H$^{13}$CN & 4--3 & 41 & 1.5$\times$10$^7$ \\
345.796 & CO & 3--2 & 33 & 1.5$\times$10$^4$ \\
346.528 & SO & 9$_8$--8$_7$ & 79 & 5.8$\times$10$^6$ \\
346.652 & SO$_2$ & 19$_{1,19}$--18$_{0,18}$ & 168 & -- \\
346.998 & H$^{13}$CO$^+$ & 4--3 & 42 & 2.9$\times$10$^6$ \\
347.331 & SiO & 8--7 & 75 & 8.4$\times$10$^6$
\enddata
\tablenotetext{a}{The critical densities $n_{cr}$ are calculated with $n_{cr} = A_{ij}/ \sum \gamma_{ij}$ under optical thin assumption \citep{2015Shirley} at a temperature of 20 K, except a temperature of 25 K for SO$_2$. The Einstein $A$ coefficients and the collision rates $\gamma_{ij}$ are obtained from the Leiden Atomic and Molecular Database \citep{2005Schoier}.}
\end{deluxetable}
\clearpage

\begin{turnpage}
\begin{deluxetable}{l@{\extracolsep{10pt}}c@{\extracolsep{8pt}}c@{\extracolsep{8pt}}c@{\extracolsep{12pt}}c@{\extracolsep{8pt}}c@{\extracolsep{8pt}}c@{\extracolsep{12pt}}c@{\extracolsep{8pt}}c@{\extracolsep{8pt}}c@{\extracolsep{12pt}}c@{\extracolsep{8pt}}c@{\extracolsep{8pt}}c@{\extracolsep{12pt}}c@{\extracolsep{8pt}}c@{\extracolsep{8pt}}c}
\tabletypesize{\scriptsize}
\tablecaption{Mapping Parameters$^a$ \label{map_table}}
\tablewidth{0pt}
\tablehead{ 
\multirow{2}{*}{Line} & \multicolumn{3}{c}{Cyg-N38} & \multicolumn{3}{c}{Cyg-N43} & \multicolumn{3}{c}{Cyg-N48} & \multicolumn{3}{c}{Cyg-N51} & \multicolumn{3}{c}{Cyg-N53} \\
\cline{2-4} \cline{5-7} \cline{8-10} \cline{11-13}  \cline{14-16}
& Beam & P.A. & rms & Beam & P.A. & rms & Beam & P.A. & rms & Beam & P.A. & rms & Beam & P.A. & rms }
\startdata
CH$_3$OH 7$_{1,7}$--6$_{1,6}$ A$^+$ & 3.20 $\times$ 2.78 & 65 & 0.13 & 4.63 $\times$ 2.83 & 55 & 0.06 & 3.35 $\times$ 3.00 & 56 & 0.13 & 4.36 $\times$ 2.80 & 59 & 0.07 & 3.48 $\times$ 3.01 & 49 & 0.27 \\
SO 8$_8$--7$_7$ & 3.14 $\times$ 2.71 & 67 & 0.12 & 3.95 $\times$ 2.62 & 55 & 0.06 & 3.24 $\times$ 2.89 & 59 & 0.14 & 3.53 $\times$ 2.37 & 58 & 0.06 & 3.38 $\times$ 2.90 & 49 & 0.28 \\
H$^{13}$CN 4--3 & 3.13 $\times$ 2.70 & 68 & 0.13 & 3.98 $\times$ 2.60 & 56 & 0.06 & 3.23 $\times$ 2.88 & 59 & 0.13 & 3.60 $\times$ 2.35 & 58 & 0.06 & 3.36 $\times$ 2.89 & 49 & 0.26 \\
CO 3--2 & 3.12 $\times$ 2.70 & 66 & 0.12 & 3.86 $\times$ 2.59 & 58 & 0.06 & 3.23 $\times$ 2.87 & 59 & 0.14 & 3.49 $\times$ 2.38 & 60 & 0.06 & 3.37 $\times$ 2.91 & 50 & 0.26 \\
SO 9$_8$--8$_7$ & 3.15 $\times$ 2.70 & 67 & 0.13 & \multicolumn{3}{c}{No data} & 3.17 $\times$ 2.89 & 58 & 0.14 & \multicolumn{3}{c}{No data} & 3.35 $\times$ 2.89 & 49 & 0.26 \\
H$^{13}$CO$^+$ 4--3 & 3.13 $\times$ 2.69 & 67 & 0.14 & \multicolumn{3}{c}{No data} & 3.17 $\times$ 2.89 & 58 & 0.14 & \multicolumn{3}{c}{No data} & 3.37 $\times$ 2.89 & 50 & 0.27 \\
SiO 8--7 & 3.12 $\times$ 2.70 & 67 & 0.13 & \multicolumn{3}{c}{No data} & 3.16 $\times$ 2.89 & 59 & 0.14 & \multicolumn{3}{c}{No data} & 3.37 $\times$ 2.90 & 50 & 0.31 
\enddata
\tablenotetext{a}{Beam in units of arcsec, P.A. in units of degree, and rms noise in units of Jy beam$^{-1}$} 
\end{deluxetable}
\end{turnpage}
\clearpage

\begin{deluxetable}{cc ccc}
\tabletypesize{\scriptsize}
\tablecaption{Spectroscopic values for H$^{13}$CN 4--3 hyperfine components}
\tablewidth{0pt}
\tablehead{Rest Freq. (GHz)  & J & F & g$_{up}$ & A$_{ij}^a$ (s$^{-1}$)}
\startdata
345.3381613 & 4--3 & 4--4 & 9  & 1.19 $\times$ 10$^{-4}$ \\ 
345.3396586 & 4--3 & 3--2 & 7  & 1.74 $\times$ 10$^{-3}$ \\
345.3397693 & 4--3 & 4--3 & 9  & 1.78 $\times$ 10$^{-3}$ \\
345.3398148 & 4--3 & 5--4 & 11 & 1.90 $\times$ 10$^{-3}$ \\
345.3401382 & 4--3 & 3--4 & 7  & 2.42 $\times$ 10$^{-6}$ \\
345.3417462 & 4--3 & 3--3 & 7  & 1.53 $\times$ 10$^{-4}$
\enddata
\tablenotetext{a}{Private communication with Dr. Laurent Pagani} 
\label{table_h13cn}
\end{deluxetable}

\begin{deluxetable}{ccccccc}
\tabletypesize{\scriptsize}
\tablecaption{Fitting parameters of velocity dispersion spectrum}
\tablewidth{0pt}
\tablehead{ 
\multirow{2}{*}{Source} & a & b & \multirow{2}{*}{n} & $V_{AD}$ & $L_{AD}$ & $B_{pos}$ \\
 & (km$^2$ s$^{-2}$) & (km$^2$ s$^{-2}$ arcsec$^{-2}$) & & (km s$^{-1}$) & (mpc) & (mG)}
\startdata
Cyg-N38 & -0.50 $\pm$ 0.17 & 0.69 $\pm$ 0.04 & 0.44 $\pm$ 0.04 & 0.31 & 4.9 & 1.9 \\
Cyg-N48 & -0.30 $\pm$ 0.12 & 0.20 $\pm$ 0.06 & 0.82 $\pm$ 0.15 & 0.34 & 17.2 & 5.1 \\
Cyg-N53 & -3.00 $\pm$ 0.42 & 0.23 $\pm$ 0.04 & 0.52 $\pm$ 0.08 & 0.84 & 1425 & 6.0 $\times$ 10$^2$ 
\enddata
\label{table_vds}
\end{deluxetable}
\clearpage

\clearpage


\clearpage

\clearpage

\clearpage

\clearpage

\clearpage

\clearpage 

\clearpage

\clearpage

\clearpage

\clearpage

\clearpage

\clearpage

\clearpage

\clearpage

\end{document}